\documentclass [reprint aps,prb,twocolumn,groupedaddress,nofootinbib,floatfix,amsmath,mssymb,prper,floatfix,twocolumn, superscriptaddress]{revtex4-2}
\usepackage{array, makecell}
\usepackage{graphicx} 
\graphicspath{{./Graphics/}}
\usepackage{graphics}
\usepackage{epsfig}
\usepackage{dcolumn}
\usepackage{amsfonts,color}
\usepackage[utf8]{inputenc}
\usepackage{bm}
\usepackage{scalerel,amssymb}

\begin{document}
\preprint{APS/123-QED}
\title{Toward a Longitudinal Multifaceted View of Remote Undergraduate Research Experiences During the COVID-19 Pandemic: The perspectives of doing research online.}
%Who says research can't be done remotely?}%: Reality of the Remote Undergraduate Research Experience During the COVID-19 Pandemic}
\author{Dina Zohrabi Alaee} \author{Benjamin M. Zwickl}
\affiliation{%
School of Physics and Astronomy, Rochester Institute of Technology, 84 Lomb Memorial Drive, Rochester, NY,14623}%
\date{\today}
\begin{abstract}
In the Summer of 2020, as COVID-19 limited in-person research opportunities and created additional barriers for many students, institutions either canceled or remotely hosted their Research Experience for Undergraduates (REU) programs. The present longitudinal qualitative phenomenographic study was designed to explore some of the tools, communities, norms, and division of labors that make up the remote research program and some of the possible limitations, challenges and outcomes of this remote experience. Within the context of undergraduate research program, the paper makes a comparison between outcomes of the remote and in-person research experience. Overall, 94 interviews were conducted with paired participants; mentees ($N$=10) and mentors ($N$=8) from six different REU programs. By drawing on Cultural-Historical Activity Theory as a framework, our study has explained the remote format has not opened a gap between mentees' research and academic objectives from the REU program and what they got at the end. All mentees reported that this experience was highly beneficial. Comparisons between the outcomes of the remote REU and the published outcomes of in-person UREs revealed many similar benefits of undergraduate research. Our study suggests that remote research programs could be considered as a means to expand access to research experiences for some students even after COVID-19 restrictions are lifted.
\end{abstract}
\maketitle

\section{Introduction}
%The COVID-19 pandemic made some undergraduates research programs transition to a remote research format. 
Undergraduate research experiences affect students' academic pathways and career preparation towards STEM by providing authentic research-based learning situations \cite{council_transforming_1999, council_improving_2002, wenzel_enhancing_2004, Kenny_2001, kuh_high-impact_2008}. A large body of literature has reported academic and psychosocial benefits of in-person UREs. For instance, academically, UREs have been reported to help students achieve a higher level of content knowledge \cite{kardash_evaluation_2000}, and career outcomes after participation in UREs \cite{lopatto_undergraduate_2007, hathaway_relationship_2002}. Psychosocially, UREs have been shown to help students increase self-confidence \cite{seymour_establishing_2004, lopatto_survey_2004, hunter_becoming_2007, laursen_undergraduate_2010}, develop communication skills \cite{kardash_evaluation_2000, hunter_becoming_2007}, improve scientific identity \cite{lopatto_undergraduate_2007,estrada_toward_2011, graham_increasing_2013}, and grow a sense of belonging to the community \cite{lopatto_undergraduate_2007, hausmann_sense_2007, dolan_toward_2009, eagan_making_2013}.

One form of UREs is the Research Experiences for Undergraduates (REU) program, which is a ten-week summer research experience funded by the US National Science Foundation. Due to the COVID-19 pandemic some REU programs transitioned to a remote format in summer 2020. Interestingly, many of the positive outcomes that are cited above are similar to the psychosocial growth observed during remote REU programs in summer 2020 \cite{zohrabi_alaee_impact_2022}. Zohrabi Alaee \textit{et al.} found that students who attended a remote research program generally experienced a greater sense of belonging, self-efficacy, and physics and researcher identity despite the absence of physical proximity to lab mates and research mentors. Their results show how different aspects of their research experiences affect their psychosocial gains development.
Despite differences in the individual goals of each research project, there are several common goals between all URE programs; they are all hope to increase retention in a STEM career pathway, promote STEM knowledge and practices, and integrate students into STEM culture \cite{national_academies_of_sciences_undergraduate_2017}. Studies on in-person UREs have found a relationship between participation in UREs and retention in STEM \cite{national_academies_of_sciences_undergraduate_2017}. However, many of those studies do not fully describe why a URE would lead to increased retention in STEM among participants. In most of these studies, data have been derived from self-reported surveys \cite{hathaway_relationship_2002, blockus_strengthening_2016}, and end-of-program formal evaluations  \cite{hathaway_relationship_2002, blockus_strengthening_2016, lopatto_undergraduate_2007, russell_benefits_2007}, while fewer studies have used multiple interviews \cite{seymour_establishing_2004, hunter_becoming_2007}. %In contrast, fewer focused on a mechanism to understand why a URE would lead to increased retention in STEM among participants.
Besides, most studies have explored the positive benefits of in-person undergraduate research experience, and very little research has yet focused on a remote research experience \cite{forrester_how_2021}.

The present study used longitudinal semi-structured interviews with mentor-mentee pairs over the goal-directed undergraduate summer research activity. We explored how students achieved their project goals remotely and how the elements of a particular project lead to the broader outcomes that are common across lots of UREs such as retention in STEM. Every week of the REU program we interviewed either mentors or mentees, and we interviewed each group one additional time a few weeks after the program finished.
% In this paper we use activity system analysis to describe the research experience and understand challenges and outcomes.

The research questions addressed in this paper are:
\begin{enumerate}
    \item How did members of the REU community use different elements of an activity system to accomplish their goal-directed activity? 
    \item What challenges were observed within the goal-directed REU activity?
    \item What outcomes of remote REU programs were observed?
    %\item How do outcomes of the remote and in-person research experience compare to each other? 
\end{enumerate}

During interviews, participants described their goals, procedures, and what they achieved by the end of the program. Providing comprehensive data on mentees' progress during and after the UREs could help us better understand the complexity of the undergraduate research experience and improve the quality of similar experiences in the future. 

%In order to encourage more undergraduate students from outside the awarded institution in academic research, the United States' National Science Foundation (NSF) established the first Research Experiences for Undergraduates (REU) program in 1980 \cite{nsf_nsfs_1990}.
%, nsf_national_2013, corley_mentoring_2013, li_integrating_2015}. %\cite{lasurias_critical_2012, yuan_reu_2013, badawy_assessment_2015, follmer_preliminary_2015, nadelson_lifes_2015, lasota_preparing_2015}. 

\section{Theoretical background}
\label{Sec: Theoretical background}
\begin{figure}[tbh]
\centering
\includegraphics [trim=150 240 480 
110,clip,width=80mm]{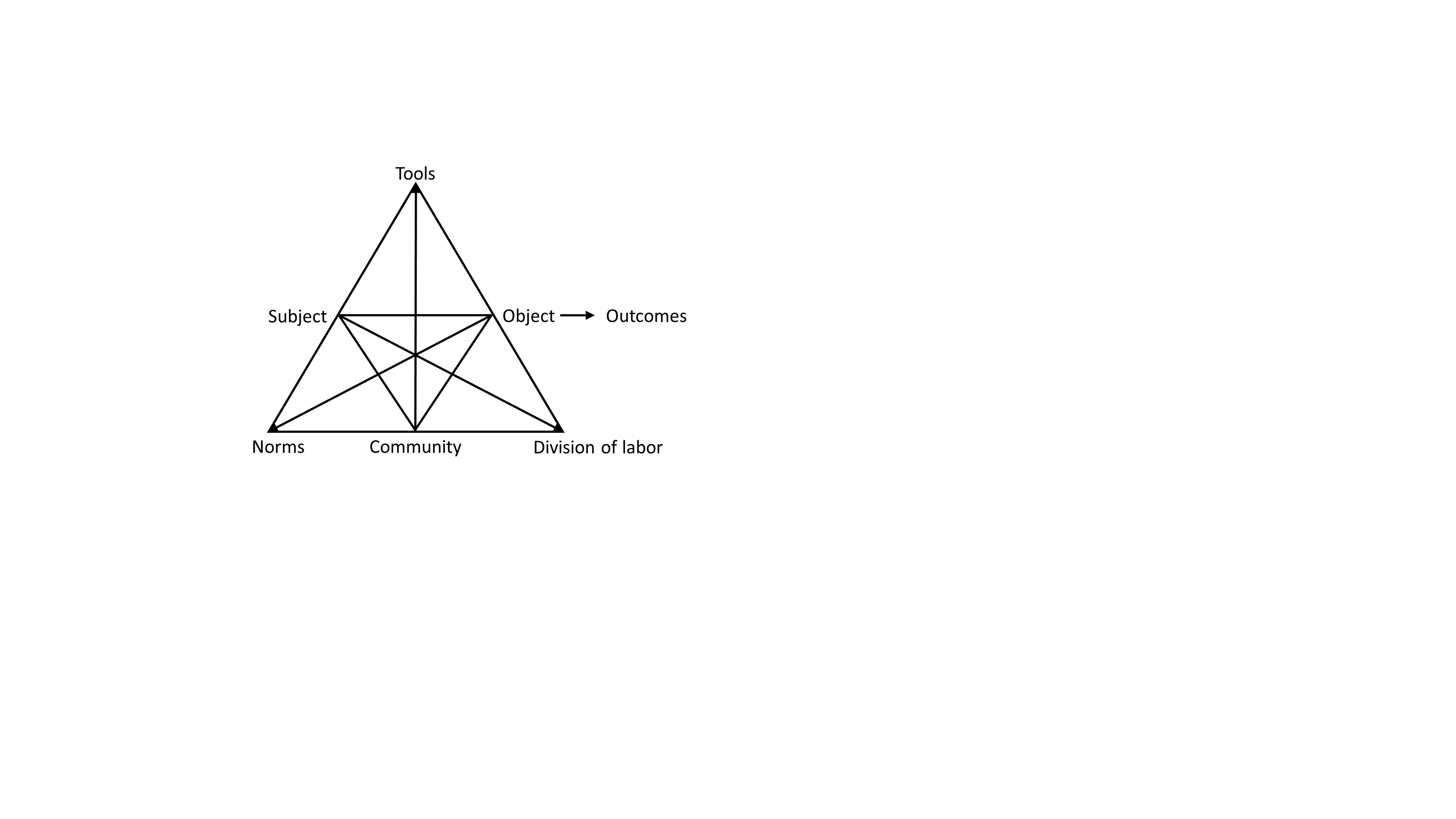}
\centering
\caption{An overview of an activity system is represented as a triangular model \cite{engestrom_learning_1987}. The Research Experiences for Undergraduates (REU) program is an activity system. \textit{Subject} refers to mentees who participated in the remote REU programs. The \textit{tools} include lab environment, software, and REU activities such as seminars. The \textit{objective} is a goal for the remote REU program. The \textit{norms} refer to embedded rules among members of the community. The \textit{division of labor} is how tasks are shared between members of the lab group. The \textit{community} refers to all members of the research lab and  REU participants. The \textit{outcome} of the REU program is the result of participation in this remote REU experience.}
\label{Fig: Fig_1}
\end{figure}

Our theoretical framework applies key ideas from Cultural-Historical Activity Theory (CHAT) \cite{leontev_problem_1974, roth_vygotskys_2007, engestrom_learning_1987} to help us identify and describe the dynamic of research practices in the remote REU setting and possible outcomes.  
CHAT is a multiple disciplines research framework that has its origin in the works of Vygotsky, Leont'ev, and Engestr\"{o}m on mental development and educational problems \cite{vygotsky_mind_1978, leontev_problem_1974, engestrom_learning_2014, engestrom_expansive_1999}. Engestr\"{o}m's third-generation cultural historical activity theory argued that every human action is goal-oriented and mediated by different components such as tools, division of labor, rules, and community (See Figure~\ref{Fig: Fig_1}).

Within the educational contexts, CHAT looks at the complex learning environment from a holistic perspective by focusing on learning as an objective which mediated by different components such as tools, division of labor, rules, and community (See activity systems triangle Figure~\ref{Fig: Fig_1}). The link between the elements indicates that they are dynamic and interact with the other components \cite{engestrom_learning_1987} to describe how students learning can be experienced in any education setting.

Because research programs are a goal-directed activity that relies upon both community and tools to make advances, we used CHAT as a framework to analyze how research was done in a remote setting. Through the lens of CHAT, the remote learning activity is described within social aspects (such as practices of the community), and cultural-historical approaches (such as identifying the tools and norms). CHAT would allow us to identify the mentees' objectives and understand the remote REU program's different components (nodes) \cite{yamagata2010activity}. 

Subject refers to an individual or group whose perspective is considered for analysis. For instance, we identify each individual mentee as a subject with particular personal input (e.g., emotions, family backgrounds). The objective is the goal that is the focus of the activity system. In our study, objectives are mentees' long-term goals for undergraduate research programs. Artifacts or tools are material, symbolic, and conceptual resources used to meditate between the student and their goals. Artifacts or tools include the physical lab environment, mathematics, software, professional development seminars, and experimental kits. The community refers to the people who have shared goals, including research mentors, lab group members, REU participants, and possibly a larger scientific community. Norms refer to the values, expectations, and guidelines for the subject to participate effectively as a community member. For example, how often and by what means should the mentee reach out to the mentor or other group members to receive help. Lastly, the division of labor describes the different roles performed by the community as they work toward a common objective. For an REU, the division of labor is how different tasks are shared between REU students and their mentors or other lab mates. Consider the example of a research mentor who wishes to help their mentees toward their research goal (objective). Within the lab group (community) the mentor introduces a new programming language to learn (tool). The mentor shares sample source code with the students (division of labour between mentee and mentor). Depending on their specific project goals, each mentee works on a different part of the code and develops that section (division of labour between mentees).

%magine a student is responsible for updating, writing or using some code in the lab. Another group member (or the PI) wrote a lot of the code and sets up part of the code for the student, but creates a more usable set of code that the student can then work with. Or maybe you can elaborate on your current example
%In short, in this study, we used CHAT to understand the remote REU educational context and possible positive outcomes of REU, such as identity, self-efficacy, sense of belonging, and persistence in STEM majors.
This paper focuses on findings from multiple interviews with undergraduate physics mentees and their mentors who participated in a remote REU program in the summer of 2020. %We discuss outcomes from remote REU without focusing on the mechanism. %of why those outcomes are shaped. 
This study is a part of a more extensive analysis of the mechanisms that explains why some of the outcomes ( See Section \ref{Sec: Outcomes rq3}) were shaped in a remote research setting and how these outcomes impact mentees' future career decision-making.

\section{Methodology}
This article explores the REU program from mentors' and mentees' perspectives. Analyzing each group helps us answer the question of how remote research was experienced in the summer of 2020. We adopted a qualitative longitudinal phenomenographic design using semi-structured interviews to collect data throughout the REU program and after it finished. This design helped us capture many features of the activity system that were not evident until the student has more time in the group (e.g., their understanding of community and norms). Our longitudinal study also pays attention to students' growth and change over the time of the program.

\subsection{Data collection}
We sent 64 physics REU program coordinators an email asking if their REU would be in a remote format in the summer of 2020. We received eight positive answers. Mentees were recruited into this study by an invitation email from their REU coordinators on behalf of us. After students volunteered to participate in our study, we contacted their mentors. Our overall sample included ten mentees and eight paired mentors from six REU programs. Demographics of all participants are shown in Table~\ref{tab:demographics-participants}. The sample of mentees was gender and ethnically mixed, while the sample of mentors was all men. We were unable to recruit women mentors. While the percentage of women among physics and astronomy faculty members has grown in recent years, in 2019 women were 19\% of physics faculty members and 23\% of astronomy faculty members \cite{AIP2019}. Since the COVID-19 pandemic started, all faculty were negatively affected \cite{AIP2022}. However, women faculty often faced greater challenges due to additional responsibilities of caregiving for other family members \cite{myers2020unequal, katz2021re}. All mentees were physics majors. In Table~\ref{tab: Project}, we briefly outline the titles and abstracts of each mentee's project, which represented a variety of research areas in physics (all participant names are pseudonyms). The students were compensated with \$20 gift card for their participation in each interview that sent weekly.

\begin{table}[thb]
\begin{tabular}{l|l|c}
\hline 
\hline 
\textbf{\makecell[l]{Mentees' \\Characteristics}} &{}& \textbf{\makecell[c]{Number\\($N$=10)}}\\
\hline 

 \makecell[l]{}	& \makecell[l]{Women}	& {4} \\
{Gender}	& \makecell[l]{Men}	& {6} \\ 
\hline 
	& \makecell[l]{White}	& {5} \\

 \makecell[l]{}	& \makecell[l]{Asian}	& {3}\\
{Ethnicity}	& \makecell[l]{Mixed}	& {2} \\\hline
 	& \makecell[l]{Rising senior}	& {7} \\ 
\makecell[l]{Year of college}	& \makecell[l]{Rising junior}	& {2} \\ 
		& \makecell[l]{Rising sophomore}	& {1} \\ \hline 

 \makecell[l]{Type of home}	& \makecell[l]{Ph.D. granting institutions}	& {4} \\
 \makecell[l]{institutions}	& \makecell[l]{Master's granting institutions}	& {2} \\
   \makecell[l]{}& \makecell[l]{Bachelor's granting institutions}	& {4} \\
\hline\hline 
\textbf{\makecell[l]{Mentors' \\Characteristics}} &{}& \textbf{\makecell[c]{Number\\($N$=8)}}\\
\hline 
 \makecell[l]{Gender}
	& \makecell[l]{Men}	& {8} \\ 

\hline 
 \makecell[l]{Type of REU}	& \makecell[l]{Doctoral Universities}	& {6} \\
\makecell[l]{institution}
		& \makecell[l]{Baccalaureate Colleges}	& {2} \\
\hline \hline 
\end{tabular}
\caption{Participants' characteristics}
\label{tab:demographics-participants}
\end{table}

%Approval of the study through our institutional Human Subjects Research Office was granted before recruitment and data collection begin. In addition, the consent form that was signed by all participants included the permission of video recording and information about the study purposes. All the participants were given the right to terminate their participation at any point. Participation incentives were offered in the form of \$20 gift cards for each interview that were sent weekly.

All participants were individually interviewed at multiple points throughout the REU program in the summer of 2020 and one time after the REU program finished (interview nine with mentees and interview ten with mentors). We video-recorded all interviews via Zoom with the permission of the interviewees. Mentees' interviews took between 60 and 90 minutes, while mentors' interviews took 30 and 45 minutes. Overall, 94 interviews were conducted. Table~\ref{tab: Protocol} shows the overall protocol content for each week of the interview, which was driven by CHAT framework. For example, the mentees (subject) were asked to describe what tools they use to achieve their objective? How does the community help them to achieve their objective? What kind of norms and rules does their research labs have that affect how they approach their objective?
\begin{table*}[h]
\caption{Projects' characteristics.}
\label{tab: Project}
\begin{tabular}{l|l|l}
\hline\hline
 
{Name}  & {Area of research} 
& {Description of project} \\
\hline 
{Andrew} & 
{Nuclear Physics} &
{Understand the efficiency of detector, learn about details of the old nuclear}\\
{} &{} & {reaction simulations, and refine the new simulation}\\

{Bruce} &
{Quantum Nonlinear}   
 & {Numerically model quantum optical devices, learn PyBoard coding, use digital}\\
{} &{optics} & {time delay and construct the circuit with equipment that shipped to his home} \\

{Caleb} & 
{Atomic Physics} 
 & {Examine at the atomic structure and different spectroscopies and make a model for}\\
 {} &{} & {specific properties}\\

{David} & 
{Acoustic} 
 & {Make a basic resonator model to learn the modeling program and then make an}\\
 {} &{} & {acoustic model of a reed instrument} \\
 
{Emma} & 
{Physics Education}
 & {She had two projects: 1-PER: science outreach and 2- Build a 3D-printed particle}\\
 {} &{and Particle Physics} & {trap and work with electronics that shipped to her home to make circuits}\\

{Frieda} & 
{High Energy Physics} 
 & {Use different models in high-energy physics to predict the probability of different}\\
 {} &{} & {decay modes in collisions} \\

{Grace} & 
{Solid State Physics} 
 & {Learn the density functional theory and model certain molecules and look at the}\\
 {} &{} & {dynamics of the system}\\

{Helen} & 
{Nuclear Physics} 
 & {Simulate the decay process of short lived isotopes, learn about different isotopes}\\
 {} &{} & {and different spectra fields, and literature half lives}\\

{Ivan} 	&  %yusen
{Nuclear Physics} 
	& {Gain knowledge about CMS and LHC and use simulations and experimental data to}\\
 {} &{} & {refine the codes for detection of the charged particles in a large collider experiment}\\

{Joshua} & 
{Nuclear Physics} 
 &  {Understand neutron mirror model and add new equations into the old code to solve}\\
 {} &{} & {problems related to nuclear physics}\\ 
\hline\hline

  \end{tabular}
\end{table*}

%\subsection{The Interview Process}
%To explore the mentors and mentees' perspectives in-depth, participants were individually interviewed at successive points over the duration of the REU program in the summer of 2020 and two times after the REU program finished (interview-nine and interview-ten). Figure~\ref{FIG-Weekly} shows the content of the protocol for each week of the interview. 
\begin{table*}[h]
\caption{Overview of interview study design and protocol content.}
\label{tab: Protocol}
\begin{tabular}{l|l}
\hline \hline

{Interview schedule and participants}  
& {Protocol Content} \\
\hline
{\textbf{\textit{Start of the REU program}}} & 
{}
\\
{Interview 1 with mentees {(Jun 1-6,2020)}} & 
{Career \textbf{objectives}, REU \textbf{objectives}, and personal impacts that guide}\\
{} & {mentees' future career decisions}\\

{Interview 2 with mentors {(Jun 8-12,2020)}} & 
{Describing the \textbf{lab community}, establishing a relationship with mentees,}\\
{}&{process of moving research lab into remote format}\\

{Interview 3 with mentees {(Jun 15-19,2020)}
} &
{Group \textbf{norms}, \textbf{rules}, and communication with other members of the lab}\\
{}&{\textbf{community}, establishing sense of belonging in remote lab setting}\\

{Interview 4 with mentees {(Jun 22-26,2020)}}  &
{Create a project concept maps as a learning \textbf{tools}, approaching research}\\
{}&{and using multiple \textbf{tools} in an remote environment}\\

{Interview 5 with mentors {(Jun 29-Jul3,2020)}}  &
{\textbf{Objectives} for REU students, social practices of science by doing research,}\\
{}&{evaluating mentees' progress}\\

{Interview 6 with mentees {(Jul 6-10,2020)}}  &
{Impact of REU experiences on mentees' psychosocial gains such as}\\
{}&{self-efficacy, sense of belonging, and identity}\\

{Interview 7 with mentees {(Jul 13-17,2020)}}  &
{Create a project concept maps as a learning \textbf{tools}, challenges during}\\
{}&{the REU program and different kind of research \textbf{tools} that mentees use}\\

{Interview 8 with mentees (Jul 20-31,2020)} &
{Factors that impact mentees' \textbf{objectives} and their future career}\\
{}&{\textbf{objectives} and their interests}\\
\\
{\textbf{\textit{End of the REU program}}} & 
{}\\
{Interview 9 with mentees (Sep 14-24,2020)} & 
{REU \textbf{Outcomes}, understanding impact of the remote research experience}\\
{}&{on mentees' future career \textbf{objectives}}\\

{Interview 10 with mentors (Sep 22-28,2020)} &
{Evaluate mentees' research \textbf{objectives} from mentors' view}\\
\hline\hline
\end{tabular}
\end{table*}
The interview over Zoom can be difficult for both interviewees and interviewer. Hence, it was important to create a trusting and relaxed atmosphere between the interviewer and the interviewees that encouraged them to share their experiences. During the interviews, many participants shared personal information about themselves, their families, and how this remote experience impacted their lives during the pandemic.

\subsection{Data analysis} 
In the context of this article, we focused in on each node in the activity system (e.g., community). The activity system as a whole wasn't the analysis unit, but rather each node of the activity system was the unit of our phenomenographic analysis to understand different ways of experiencing REU in a remote setting.  
%In the context of our study, The smallest unit of phenomenographic analysis is a way of experiencing REU in a remote setting.

Each interview was recorded and auto-transcribed with Zoom for analysis. After the interviews were completed, the transcripts were cleaned to fix errors and punctuation. The transcripts became the focus of our phenomenographic analysis. We used the qualitative phenomenographic analytic method to understand the various aspects of the remote REU phenomena in terms of variations in mentors' and mentees' experiences \cite{Marton1997, marton2000structure}. According to Marton and Booth \cite{Marton1997}, ``Phenomenography is focused on the ways of experiencing different phenomena, ways of seeing them, knowing about them and having skills related to them.'' It is important to note that there are no inherently right or wrong ways of experiencing it. The main goal of phenomenographic methods is to gain insights into a deeper understanding of the remote REU experience. %In this Phenomenographic study, we focused on a small number of participants and identified a number of qualitatively different ways in which they experience and perceive the REU phenomenon. 
CHAT was used to design the interview protocols so the data would span a wide range of social, cognitive, and cultural aspects of the experience. CHAT guided data collection and analysis. One of the most significant issues for our large data set was how to make it manageable. We had 94 interviews.

\subparagraph{Analysis needed for Research question 1.}
The analysis process is divided into multiple steps, including data immersion, primary analysis generating priori codes, searching for sub-codes, and defining and naming sub-codes. %We then employed the results of objectives analysis and make a comparison between those and the outcomes of in-person undergraduate research to answer research question 3.
The process of data immersion started with transcribing, cleaning, and listening to the interviews repeatedly. Analysis of the transcripts was executed using Dedoose software \cite{noauthor_dedoose_2018}. In the primary analysis (division into CHAT categories), we examined our data based on CHAT nodes (Subject, tools, community, norms, division of labor, objectives) and applied these priori codes to each excerpt. This primary analysis created a collection of quotes under each CHAT node (e.g., community). Overall, a total of 1433 segments were coded. Norms and division of labor had the least excerpts, while community, tools, and objectives had the most excerpts. Identifying instances of a division of labor and norms within the interviews was challenging. Perhaps this is because the remote research experience led to more isolated work, which lacked a common goal within the lab community and lacked regular face-to-face interactions among the community members. 

The second stage of the analysis (variation within each category) started with a rigorous process of going over every excerpt under each CHAT node and searching for significant sub-codes. The main aim of phenomenographic methods is to gain insights into a deeper understanding of mentors' and mentees' perspectives about their research experience. In other words, its aim is to discover the qualitatively different ways in which participants experience REU through the CHAT framework and elicit the variations within each node. 

%Then, we were ready to begin the rigorous process of going over every excerpt under each CHAT node and searching for significant sub-codes to elicit the variations within each node. 
%The coding process in this study consists of multiple steps, including data immersion, generating initial codes, searching for sub-codes, and defining and naming sub-codes. We then employed the results of objectives analysis and make a comparison between those and the outcomes of in-person undergraduate research to answer research question 3.
\subparagraph{Analysis needed for Research question 2.}
To address research question 2, we needed to identify challenges during the REU experience. Challenges were disruptions in the REU activity system and can provide opportunities for developing an improved version of an online REU. After analyzing data for the RQ1, we noticed some participants talked about interruptions and challenges within their research progress. We noticed tensions could occur both within one element of the activity system, such as a lack of community, or between components of the activity system, such as a lack of connection between tools and objectives. Once challenges have been identified, we clustered together all quotes that talked about similar challenges under specific sub-code through this process. In addition, we used the CHAT framework as a lens to examine the interactions between the nodes that participants mentioned during their research experience. For example, on several occasions, mentees mentioned that the lack of REU community could impact their goals for the REU program, which was to build a connection with broader scientific community. %We do not aim to construct an activity system for each participant. Instead, we sought to identify the various elements within each node to demonstrate a holistic view of what happened in a remote research format and some of its significant outcomes. 

Given this context of the methodology of our study, herein are some of the specific challenges we faced. The first one was the coding process. Since coding all 94 longitudinal transcripts was overwhelming and broad, analyzing the data and grasping the connection between the nodes was challenging. Due to the nature of a remote research experience and a lack of face-to-face interactions among the research community, %%%Additionally, it was hard to identify instances of division of labor within the interviews. Perhaps this is because the remote research experience led to more isolated work, which lacked of a common goal with their community. 

\subparagraph{Analysis needed for Research question 3.}
We used the same analysis process as RQ1 to address RQ3. We coded the transcript through the outcomes node. Then, the entire interview transcripts were re-read and searched for related sub-codes. Finally, the results from interview analysis were compared to the previous findings around outcomes of in-person undergraduate research to answer research question 3. 

%\subsection{Inter-rater reliability (IRR)}
%We used a phenomenographic approach to analyze each interview separately. The analyses of all the quotes lead to three research questions (RQ 1,2, and 3) that are discussed in this study. The major categories were collapsed into several subcategories. After several more cycles of analysis and refinement of our data sets and code book, we reached a stable point when our categories were fixed. We invited an additional researcher to separately code \text{10 \%} of our data set which was randomly selected. We found that the majority of our disagreements were in RQ1 (with \text{77 \%} agreement), between mentees' goals and mentors' goals. To address this issue, we provided more context for each quote. The inter-rater reliability for RQ2 and RQ3 were \text {98 \%} and \text {89 \%}, respectively before discussion and \text{100 \%} after discussion. The discussion process included explaining, refining, and finalizing the code book.

\section{Results on research question 1}
\textbf{\textit{How did members of the REU community use different elements of an activity system to accomplish their goal-directed activity? }}
In this section, we aim to describe all the different elements of the CHAT framework within a complex remote undergraduate research setting. There is a subsection related to each node of the CHAT framework, and under each node we include phenomenographic themes. 
%starting with objectives. Besides, within each subsection, we present the phenomenographic themes
%This paper focuses on results from interviews with the mentees ($N$=10) and mentors ($N$=8) at six U.S. institutions who were interviewed several times throughout the remote REU program and after it finished. By positioning the study within a CHAT theory, we tried to understand and describe all the elements working and connecting in a complex remote undergraduate research setting. 
\subsection{Objectives outlined by mentees}
\label{Sec: Objectives}

\begin{figure}[tbh]
\centering
\includegraphics [trim=305 90 230 
70,clip,width=85mm]{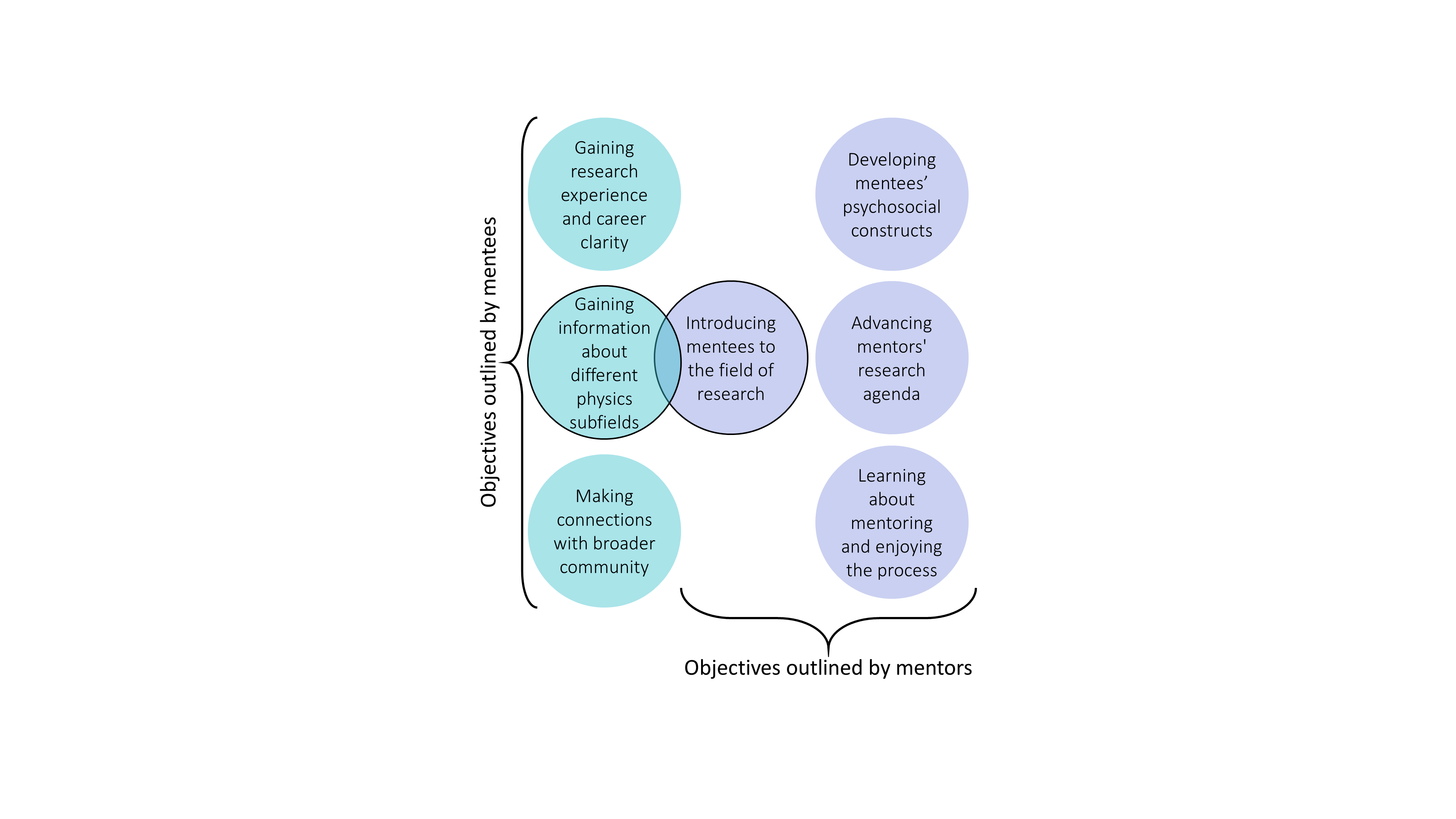}
\caption{Objectives are represented by a circle. Overlap between mentees' objectives for the REU program and objectives outlined by mentors.}
\label{Fig: Fig_2}
\end{figure}

In this study, objectives and personal goals for mentees around attending the remote research program fell under the following three categories as shown in Fig.~\ref{Fig: Fig_2}~(blue bubbles): gaining research experience and career clarity, gaining information about different physics subfields, and making connections with a broader community. 

\subparagraph{Gaining research experience and career clarity.}
%Adam, Brian , Olivia, Shuiji, Anthony, Amanda, Katelynn (narrow down my career options)%(Future career decision making) Adam, Yusen, Shuaji, Brian, Anthony, Bryce, katelynn, amanda, Olivia
This category (See~Fig. \ref{Fig: Fig_2}~(Top blue bubble)) was the most frequent goal presented in the data, meaning that it was the main reason for most mentees ($N=9$) joined the REU program. For instance, Caleb described his goal as wanting to ``See what a full-time job doing research such as this would be like and see if that would be something I am interested in doing for a career.'' For Helen, who was from a small institution, the goal of participating in the remote REU program was hearing more stories about STEM career trajectories. She said, ``Since I go to a small school and do not have that many STEM majors, I wanted to...hear what [other people] wanted to do with their physics degree and what path they want to choose.'' 
 
\subparagraph{Gaining information about different physics subfields.}
%Yusen, Anthony, Amanda, Katelynn, Adam, Olivia
Six mentees explained that they expected to gain new information about a new research field to help them increase their understanding of the topic and to narrow their future career options. For instance, David said, ``The professor I am working with is doing research in musical acoustics. %Since I am studying music as well as physics, I thought that it would be really interesting. 
I am hoping to get a better idea. So whether I want to go into industry or if I want to go straight into graduate school after this and get a better feel for the acoustics and what I might be doing in graduate school as well.'' This example fits both goals of gaining information around subfields of physics, gaining research experience, and career clarity.

\subparagraph{Making connections with broader community.}
%Olivia, Brian, Bryce, Mckeana
More than half of mentees ($N=6$) stated that they would like to make professional connections, including meeting more physics faculty and other REU students and collaborating with different people in the field (See~Fig. \ref{Fig: Fig_2}~(bottom blue bubble)). Helen believed this remote experience could give her a chance ``to meet more people interested in physics.'' However, part of this goal of meeting more REU students was not well met because of the remote work environment. (See Section \ref{Sec: Challenges-Lack of REU community})
%Some mentees expressed interest in later receiving a strong letter of recommendation. 
In addition to getting a good research experience, Bruce said, ``I want to get a good letter of recommendation if possible.'' 
% letter rec anthony, mckeana
\subsection{Objectives outlined by mentors}
Mentors outlined several goals for the remote REU program. Some of the goals were mentor-focused, such as learning about mentoring and advancing their research agenda. Other goals were focused on the REU students, such as introducing mentees to the research field and helping them to feel they are a part of the community. Figure~\ref{Fig: Fig_2}~(purple bubbles) shows the alignment of mentor and mentee objectives. 
%Only one mentor talked about a departmental obligation as a motivation to participate in the REU program. 
\subparagraph{Introducing mentees to the field of research.}
In addition, our data elucidated that mentors' execution of their goals benefited mentees by introducing them to the field, showing them more realistic views on what research looks like, and providing them with educational and professional opportunities. These goals are very close to the mentees' objectives that we earlier discussed in Figure \ref{Fig: Fig_2}~(Middle blue bubble). Mentors tried to facilitate the research process and align mentees' goals with their own goals and values. In addition, mentors focus on mentees' future career possibilities by providing them educational and professional support, such as sharing their personal academic experiences and introducing them to the field.

\subparagraph{Developing mentees' psychosocial constructs.}
Some mentors also focused on helping mentees develop their psychosocial constructs, such as a sense of belonging, self-efficacy, and science identity. For instance, one mentor explained his goal is promoting his mentees' science identity by introducing the mentee to the broader community and providing opportunities to present at conferences. He said, ``I believe very strongly that the purpose has to be the education of the student. It is not the new knowledge that you generate, although we publish actively and we give talks. Then we produce new knowledge. But the main reason for doing this is to help students become scientists. So that is what I do.'' 

\subparagraph{Advancing mentors' research agenda.}
In addition, to introduce mentees to the field and provide them with authentic research and learning experiences, some mentors talked about personal research factors that encourage them to mentor undergraduate students. For instance, David's mentor said, ``My personal [mentoring] experience has been a big benefit to my research; that's where I get...my really best students.  %because the advantage I have for the rue program is there are students looking for summer research programs in acoustics.  Which they're not very many.  And of course.
...I often, not always but often will get a student, in fact, interested in studying acoustics and going to Acoustical Society meetings to present because they're interested in graduate school. So that it's been an REU program has been a real benefit to me.''%..I think it really helps with graduate schools [too], where you can get a reference from somebody like me. This person actually worked with him on research; it's going to be a much more meaningful reference than somebody writing a reference where he took my course. He got 93\%, and he's one of the top 10\% of the class.''

\subparagraph{Learning about mentoring and enjoying the process.}
There is some evidence that mentors generally enjoyed working with REU students. For example, one of the senior mentors stated, ``I enjoy working with these brilliant young students. They can be from anywhere. So I work with people from all over the country. Maybe all over the world, from other countries, so that is really fun because the students are very good and energetic and eager to learn.'' Another mentor from a PhD granting institution commented, ``A mentor should learn something about themselves. A successful experience for a mentor is learning more about mentoring and how to have better relationships with the people you are doing research with so that you have more productive research in the future.'' 
\subsection{Mediating tools discussed by mentees}
Actions in the CHAT framework are goal-oriented and those actions are facilitated by tools. These material, symbolic, and conceptual tools in CHAT either mediate mentees' efforts to do research successfully in the remote research environment. The remote REU program was a new experience for all the participants. Hence we first asked mentors how they used new tools to switch their research lab to the remote format. Since experimental data collection was impossible, mentors emphasized other aspects of research beyond experimental data collection.

%Emma's mentor discussed his thought process around setting the REU goals. He said, ``What was the original goal of the REU program? Can that goal be met in a remote format? If it cannot, we need a new program.'' 
For some mentors, the program's goal was to give mentees an immersive experience of working in the lab, which was not possible during the pandemic. From that point of view, they needed to design their program around new goals. For instance, Andrew's mentor stated, ``We wonder what can they do from home, and what they can do from home is programming—programming what? Well, in nuclear physics, what we do is analyze data, and we felt a little weird to dump data on them without context. So, the alternative is to simulate. We realized that regardless of what they do in their scientific career, how they will think about that program will positively impact them. So that is why we chose to do [simulations].''

In contrast, three mentors did not change much about their objectives in this new remote format, likely due to their research nature and the technical aspects. For example, David's lab group work was primarily computational. Their group was not impacted by the transition to a remote work format. David's mentor explained, ``It is similar to what we would do if we were in the same room every day. Of course, we are not all in the same room, even when we were in the same room, especially for students who do the kind of thing that the mentee is doing. If he is working, he is sitting at a computer here...with my student last year, when I was in the same room, I would be sitting at the computer most of the day, and I might drop in. He might have a comment or question occasionally. You go for actually pretty long periods without any direct interaction.''

\begin{figure}[tbh]
\centering
\includegraphics [trim=295 110 330
100,clip,width=82mm]{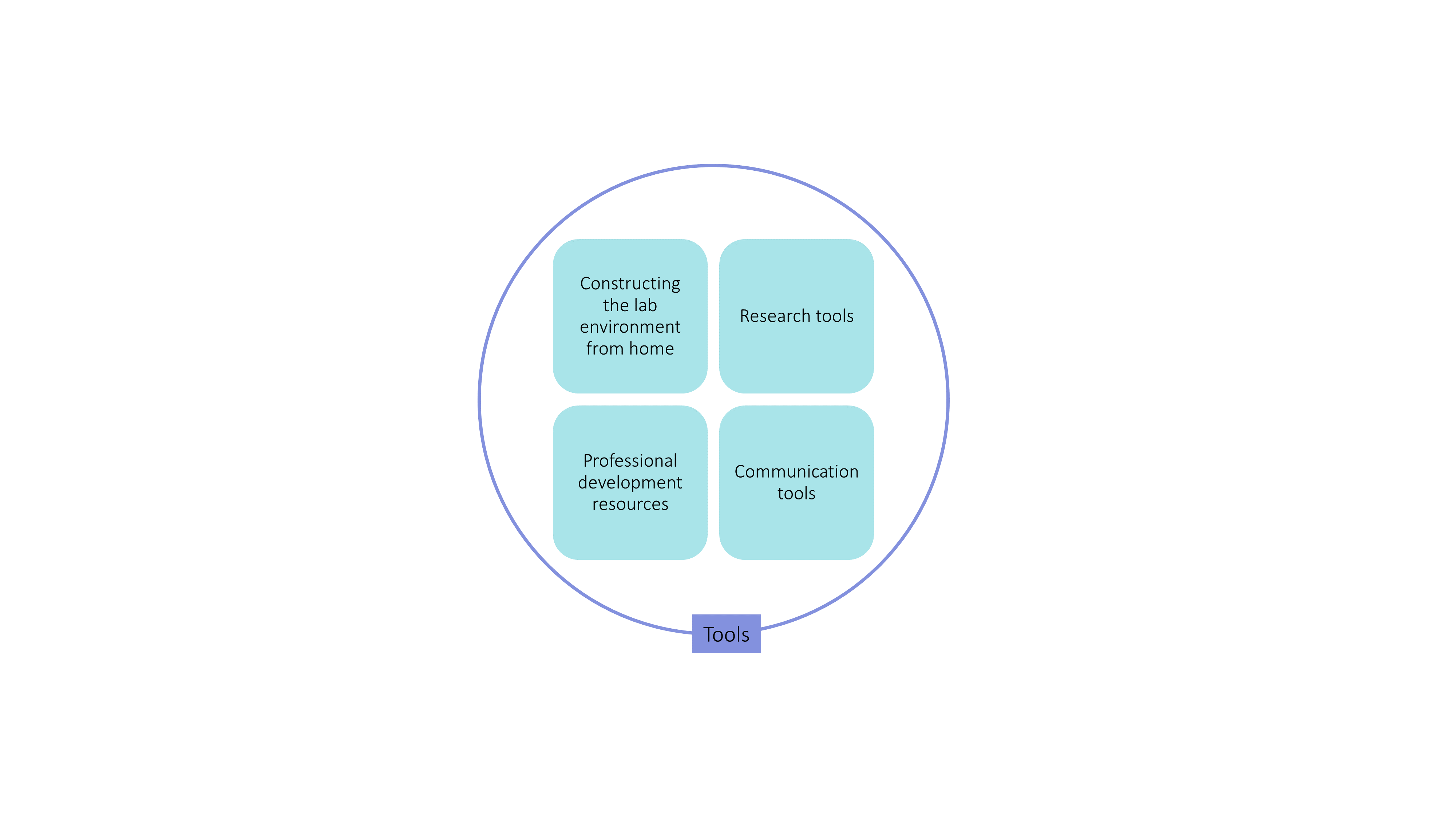}
\caption{The four tools used by mentees in the remote REU program.}\label{Fig: Fig_3}\end{figure}

Then, we asked mentees how they used tools to do research in a remote setting. As a result of our analysis (Fig.~\ref{Fig: Fig_3}), four most repeated tools categories were identified among mentees' responses, including constructing the
lab environment from home, learning tools, professional development resources, and communication tools.

\subsubsection{Constructing the lab environment from home}
\label{Sec: Constructing the lab environment from home}
Faced with the COVID-19 pandemic, one fundamental change was switching from a physical laboratory environment to a remote lab environment. Seven mentees lived with their family and had access to separate bedrooms, a basement, or other relatives' houses. One mentee lived with her partner and worked from the basement, and two Chinese students lived in dorms with other roommates. The home work space is filled with various sources of distraction such as hanging out with siblings, using technologies such as TV and video games, texting friends, and watching videos on YouTube while working. 
\subparagraph{Working at home features digital distraction.}
Working from home created a new range of digital distractions. Andrew, who had an online class experience, said, ``When I am in front of a computer, it is very easy to get distracted.'' He continued, ``I am very much a person who needs to be separated from my technology when learning...being at the computer when a game is a click away or a YouTube videos that always click like a major distraction.'' 

Caleb, who moved his workspace to the basement, said, ``You are at your house, so it is too much distraction. If you are in a lab, all you really have is to work. But, here at your house, the TV is right there. The fridge is upstairs, and my Xbox is right behind me.''
%Caleb said, in order not to get distracted and staying motivated, ``I have like my own personal workspace setup here at home in my office in the basement. The the hardest part is having it be in your own house where you eat, sleep, all that stuff. I just have a separate area set aside. It feels as real as it can.''
\subparagraph{Establishing designated spaces for work.}
The physical setting where several participants worked was transformed to provide a dedicate space for research. Andrew said he used a ``do not knock'' sign while working. For Bruce who lived with five other family members at home is a little bit different. He said, ``I tried to split the room in half physically. So half the room is the bedroom, and the other half is an office. When I am working, I am pretty much confined to the room. If I want to do readings, I can go outside. We have a pretty large backyard, which it has been quite pleasant, so I can go out there and do some readings.'' He also mentioned working from home allowed him to work at night because his home is ``quite hot during the day'' and that he benefited had quality time with his family.  

Emma set up her office in the storage closet in the basement to have a separate workspace. She said, ``Everyone thought it was really weird. But, I think to me, when I walk in, it is like, all right! I am doing the work right now. That is really helpful to me. As opposed to having a huge room with access to all kinds of things, I shut myself in here. There is only one door that leads out, and I'm here to work, and it feels like this is a work environment.''
  
\subparagraph{Blurred lines between personal life and research work.}
Working from home around family can both mean dealing with negatives, such as digital distractions, and positives, such as re-balancing family life. For someone like Bruce, who described himself as an ``introverted person'' and said he preferred not to hang out with anyone, working from home was a beneficial experience. He said, ``This is the first time in the last three years that I have spent a meaningfully long time with my family...I have been taking this quarantine time, and I have not been really focusing on my growth has been around my academics or my research, but it is really more around how I interact with my family. So, maybe I have changed as a researcher, but it has totally been overshadowed by how my relationship with my family has been changing.'' Notably, he noticed considerable growth in his physics and researcher identity and his future academic decisions by the end of the REU program. 

However, for David, family life was more of a distraction. He said, ``It's always more difficult to focus on the work when I'm at home, rather than at school or somewhere else. There's just all the time I spend with my family and just being around them.'' Similarly, Caleb said, ``There are many distractions when you are in your own house. Like the buddy texts you to hang out...so, that was hard at first, then I just got used to that and went from there.'' 
%Besides, one mentee talked about how he would hope to travel to a new institution and experience a new lifestyle.
\subparagraph{Working from home can boost flexibility.}
For some other mentees working from home was a valuable experience, and they enjoyed it. For Grace, who was doing computation, working from home was ``doable on online format''. She said, ``I enjoy being able to lay down sometimes; I do not like sitting in the same spot. Sometimes I have the desk, but sometimes I will spread out on the floor. I will work on the floor for a while, stretch out, and move back up. So, I like having that ability. ''Frieda did not have her own personal workspace. She said, ``Since I have started [the REU program], I have stayed at home, at my aunt's house for a week, and at my grandparent's house for a week where I currently am, where the dog has her head in my lap. I move around to different rooms...I do not like being in the same place all the time. I feel that people respect my right to work like they are quiet when I am working.''

\subparagraph{Working remotely from home leads to thriving while doing research.}
One aspect of working remotely was how it made students think deeply about the kind of work they want to do in their future careers and how to achieve a good work-life balance. For example, David said, ``I feel the remote aspect of it was very interesting and helped me to understand a little bit more about my personal connection to the work that I was doing since I wasn't at a place, doing the research. It was instead like at home; I was doing the research. So that helped me to figure out my priorities and just how much I enjoy the research and connection with the other things that I do it normally.'' 

When the interviewer asked mentees for their overall feelings every morning when they sat behind their desk to work from home, everyone responded positively, even though working from home meant dealing with distractions for some of them. For Emma, it was ``Pretty good, because I have my routine. I get up in the morning, and I have my kettle. I set up a nice spot. My office is in the basement. There is a sink-in, an outlet, and stuff. I set up my electric kettle there, and I have instant oatmeal stuff and all my tea. In the morning, I will turn the kettle on like 10 minutes before I need to do something, make my oatmeal, and make a pot of tea. So, I have the little morning routine that I do. It feels nice...I feel excited about what I am doing and what I am learning. So I have enjoyed it.'' In order to make working remotely from home more successful, students need specific resources such as a separate workspace, fast internet, and a personal computer.

\subsubsection{Research tools}
The section describes various research tools participants usually used to pursue their research objectives in a remote research environment.
\subparagraph{Use programming languages and software packages.}
All mentees used digital tools to pursue their project objectives. The most frequently identified tools across all mentees throughout the REU program were the programming languages and software packages. Bruce said, ``There is already some code written by someone else in MATLAB, and the idea is for me to be able to move that to Python.'' 
\subparagraph{It is important to keep a good notebook.}
Three mentees talked about how their mentor encouraged them from the beginning to practice science by keeping a good lab notebook, taking notes every day, and tracking their questions and their thoughts. For instance, Grace was documenting her progress in a daily lab notebook. She said, ``We do not have labs in our houses, so any changes we have made to the code or outcomes that we have had for the day, anything we have read, and anything we have questions on the kind of as a log in the lab notebook.'' 
\subparagraph{At-home experimental projects.}
Working with equipment is one of the important elements of any in-person research program that would typically be missing in a remote lab setting. Two mentors provided an experiment kit delivered to the mentees' homes (Bruce and Emma). 
%Bruce's mentor said, ``Bruce is quite interested and gifted in terms of numerical modeling and theory... We were able to send him all the instrumentation to his home. We could do that with NSF funding. He was able to put together an optic electronic oscillator, model it, and solve the delay differential equations he used to model it... I think he did a really good job of both learning how to model these kinds of systems and actually to build a system.'' Interestingly, 
Using the kit allowed Bruce to work with the equipment, making him feel like ``I got a real research experience. I think part of that was because I was actually able to perform an experimental project.''

Using a 3D printer has changed the whole learning dynamic for Emma. She was working on a 3D-printed ion trap that was broken. She said, ``This is a project that he actually did with some other students last year and within the last couple of years, but they couldn't get it to work. My job first would be to try to get the conductive filament working with the 3D printer and then to try to make the ion trap.'' Emma was able to go to campus to learn how to use 3D printers because she lived in the same town as the REU institution. She could ``read procedures'' and ``work with electronics and create circuits'' to fix them. She said, ``I just kind of went through each part to look at it and figure out that parts need to be conductive and that they were no longer conductive. So we need to reprint those parts and coat them with conductive paint...using a 3D printer was [much] fun.'' She worked remotely most of the time but she went to campus  to use 3D printer and her mentor let her ``Print [anything she wants to practice] as long as it does not take up a whole bunch of supplies. I printed my partner and I little cat rings.''%... [Moreover], by the end of it, I was pretty well versed in how to use the 3D printer... You would think it was a straightforward thing, but using 3D printers is not straightforward.''

\subsubsection{Professional development resources}
All REU programs organized a series of workshops and professional development activities outside of the research project, which were held every week by faculty mentors. For instance, some programs had professional development seminars (e.g., graduate school, future careers) or other workshops (e.g., research ethics, Python programming). These remote professional development sessions informed mentees about different physics subfields and about different career paths, which were aligned with the goals that mentees listed earlier in section~\ref{Sec: Objectives}. 

For instance, Grace attended a couple of professional development workshops and said, ``It was pretty good. I feel like there was a lot of information on the GRE, what we're going to have to do, and when they're available when you can take them...what kinds of fields like what GRE different fields are looking for. So like you're going into astronomy, what they would like to see versus someone going into  Physical Chemistry. I felt that was very informative. I did not know where to look for GRE stuff before, other than just googling...then we also talked about applying to graduate schools, and the application process, like how, do you narrow down your graduate school and how do you find someone you would like to work with.'' Most mentees had a positive experience about these professional resources. For instance,  Emma had two science ethics courses, and she said, ``I have never had to take a science ethics course before...it was all the stuff you would expect, but it was cool to hear physics professors talk about things with specific physics examples. That is not something that was ever offered or required for me. That is fun.'' However, Bruce felt a weak community outside of his research lab. He stated that they had a series of lectures on nonlinear dynamics given by different faculty. He said, ``That entire time, the REUs don't really interact with each other because we're listening to the lecture.'' These activities imparted knowledge, but did not support community. Zoom had limited options for interactive meetings, and during the early months of the COVID-19 pandemic some faculty may have benefited from more training on hosting virtual meetings.

%\textbf{Lack of BG knowledge:}
%Shuaijie Li: During the reu program. You know, there are actually some frustrating period
%Dina Zohrabi Alaee: at the beginning of the program. 
%Shuaijie Li: Yeah, thats part of the frustrating time. Yeah. The reason is that I need to learn a lot of basic background knowledge. So that's hard. One of the hard times. The other  maybe some segments of the progressing after receiving some type of progresses you know when I got a  anticipated results. I'll be very happy, but after that new difficulties and problems need to be solved again. So this time I'll change my state from excited state to another state, you know, this might be frustrated too. My challenges maybe I found no other challenges just about the program. Yeah, this is the most important part of the reu program

\subsubsection{Lab Groups utilized a variety of communication tools}
Successful research experience in any environment requires active communication with research mentors. However, in remote research space, establishing a clear relationship and open communication is a key factor of a good research experience \cite{zohrabi_alaee_impact_2022}. Effective communication such as active listening and asking questions take more effort in a remote setting, when there are no face-to-face interactions and members of the lab communicating online by Zoom, email, and text. It is more possible to increase the chance of miscommunication in such environment. All lab group communities used Zoom and email as primary platforms to share their ideas and their project results remotely. Ivan's mentor said, ``I have switched to Zoom in the last year because Zoom has become very popular, and it is nice that you do not have to register in order to join meetings.'' Caleb's mentor stated, ``Sending emails so slow and difficult. It's difficult to express yourself in emails...but Zoom is a very good thing actually for doing research.''
 
In addition, some groups used other instant communication apps in addition to Zoom, such as Slack, WeChat, Skype, etc. Grace stated that her mentor likes Zoom, but, ``The other director for the program likes WebEx. So, we go back and forth.'' Joshua's mentor said, ``We use Zoom. But, we also use WeChat. Because sometimes Zoom does not work very well, we lost connection.'' Two mentors used Slack to provide their students with rapid feedback. For instance, Andrew's mentor said, `` I am not checking his own notebook. What we have is that results are posted on Slack. So results are public for the group. I provide feedback, and I provide how to present data, ...I suggest change and things like that.'' 
%Slack: Lannon, Couder, 
Emma's mentor gave her his phone number. She said, ``He says to text him if any immediate thing happens.''

Four lab groups documented their research progress in Google Drive, one group used Dropbox, while the others communicated by email and verbally over Zoom. Grace said, ``We do Zoom, email, and then Dropbox for all documentation. [My mentor] put any sort of PowerPoint or presentation in there as well.'' Joshua's mentor described challenges when collaborating with Joshua who was living in China, said, ``We tried three different ways already for sharing files. I used Box.com and Google Drive, which China blocks Google stuff. He did not get any access. So try the Box.com could not have access, either. Then, I have a website, and I can upload some files there, so I plan to put my files there.''
% google drive: tan, lannon , cotingham, kilburn 
Effective feedback during the REU program can help mentees adapt to the remote research experience. Mentees' ($N=7$) responses show feedback is valued. Their quotes included ``my mentor ask me to look at the big question'', ``provide me different resources'', and ``suggestions of different approaches to go in different directions to go, rather than walking me through the entire process.'' For instance, Andrew said, ``[My mentor] is always on Slack, which is what we're using to communicate. So if we have a question at any time. We can just get right to [him].''
%Throughout the mentees comments, they valued ($N$=7) the opportunity to get feedback from their mentors. For example, quotes included ``ask me to look at the big question'', ``provide different resources'', and ``suggestions of different approaches to go in different directions to go, rather than walking me through the entire process.'' There was only one mentee who mentioned her advisor was not responding to her questions fast enough because he was busy with his family. Helen stated that, `` I just emailed them, the other day, but then you look on your email, it was actually like a week ago. I think days go by fast and you kind of lose track when you're not seeing each other.''
%%%%%%%%%%%%%%%%%%%%%norms

%One group consisted of a mentor who grew up in China and a mentee from a Chinese university. The mentor stated that he could explain some physics concepts in Chinese instead of English to help his Chinese mentee understand complex physics concepts. 

%Overall, most mentees ($N$=6) had meetings more than once a week, some ($N$=3) met their mentor only once a week, and only Ivan met less frequently than once per week and mostly communicated through email. He said, ``We do not have virtual meetings. We send emails, and [my mentor] sends emails back. Maybe my mentor can watch what I do from the internet; they know what I have done.''% I just tell him that I have done this and that, and he gives me another task. Those tasks are small and I think they are very simple. I think if I had to do this without the help from my mentor and I think I can do it by myself.''

Overall, the participants had a positive feeling about using online communication tools. 
For instance, Andrew felt pretty active and comfortable in discussing with other REU students remotely. He said, ``I tried to contact them multiple times a week because obviously, they are online too...the sort of dynamic we had over discussing things back and forth throughout the day kind of made me more sociable. I was not worried about getting on a Zoom call with them, talking to them, discussing things, and helping someone if they had a code problem. So I think it advanced me.'' 

However, virtual interactions may be more challenging for some students. Bruce said, ``I am more comfortable asking a question in-person than emailing a question. Personally, I feel like I am not that great via email.'' Helen, who had a mentor with two kids, said, ``We meet once as a big group and then me and him just email. I think that would be helpful for him. I also like met individually, but we have not really.'' She continued, ``Somehow, take like a day'' for her mentor to respond to her questions via email, ``So it is kind of challenging...it's kind of tough when you're stuck. And then you finally reach out and you're waiting for their response for a day and you I can't really do much more until you are able to connect with them.''

\subsection{Norms explored by mentees in each research lab}
\label{SEC: Norms}
%One special type of behavior are norms around communication and how members of the community reach out to the community. 
%During the remote situation, it is more crucial for REU coordinators and mentors to provide clear expectations for communication and for working with group members since there is no opportunity to learn norms by observing the behaviors of others. %For instance, in an in-person REU program, a lab community could establish specific approaches for help-seeking; expect REU mentees to reach out to graduate students first. However, in the remote setting, usually mentors were the first approachable persons who would provide feedback. 

%Mentors can help mentees by making them feel included in the lab's culture and answering their day to day questions and concerns. For instance, David said, ``I feel relatively comfortable with sharing my ideas. Usually, during our one-on-one daily meetings, I will show or give a progress update to what I have done and what I have been able to accomplish, so I will get some feedback pretty regularly.''

In a typical in-person research lab, norms are often so routine and embedded in a community culture that members of that community are even unaware of norms and how they affect their behaviors. However, in a remote research lab, norms are even more hidden. Those that are more explicit tend to be shaped through text-based communication programs and remote interactions over Zoom. These hidden norms create challenges for new group members who don't realize there is some unwritten rule or expectation. In such a space with a lack of visual and auditory cues, it is important to establish expectations around communicating, mentoring, learning, and researching. For instance,  Frieda's mentor explained that he told his students that ``You're not getting positive feedback all the time...we talked about that very early on in the program because some students think they're going to be told every day,`you're doing a great job', pat on the back type of thing.'' He told them, ``In this field, you're constantly told what you're doing wrong, not what you're doing right...and if you're not getting yelled at. You're doing a good job.''  

\subparagraph{Regularly scheduled meetings.}
The idea that each learning community has its way of communicating is noteworthy because each group has a strong cultural aspect. For instance, one group had one meeting every week, while the other had regular 5-minutes check-ins meetings over Zoom every day. %Some groups preferred emails instead of Slack messages. 
Having multiple weekly meetings was one of the ``productive '' norms among many REU groups, in a remote lab setting. Weekly check-ins were vital to support mentees' research goals and help them develop more open communication with their mentors. For instance, Caleb's said, ``In the Zoom meetings he has set up twice a week, you are expected to have not a presentation but stuff to show him and tell him. He can see if you are on the right track. That really encourages you to work on it as it gives you goals and kind of steppingstones along the way. So, it is not just like I am sitting here, and way down the road is the end goal. So you have stepping stones along the way, which I think kind of breaks it up nicely and make sure you are doing the right thing.'' On the other hand, Ivan was in a very different time zone from his mentor. He said, ``We do not have virtual meetings. We send emails, and [my mentor] sends emails back. Maybe my mentor can watch what I do from the internet; they know what I have done.''

Overall, most mentees ($N$=6) had meetings more than once a week, some ($N$=3) met their mentor only once a week, and only Ivan met less frequently than once per week and mostly communicated through email.

\subparagraph{Misunderstanding group norms.}
%Frieda described her mentor as a ``nice and friendly'' person. She continues working with his mentor after the REU program finished and looking at graduate school applications. Her mentor said, ``We met for an hour, every day...I tried to keep things light in the sense that we would often start meetings just joking around about stuff, we show each other's pets to each other, that was kind of building a little rapport.'' However, 
Ivan's mentor's friendly behavior differed from what Ivan, an international student, expected. He said his ``research lab dynamic is not very serious...they give me some jokes about remote working...the [mentor] and the host of the REU program is very humorous, and actually, the [mentor] is very serious with other people but didn't serious with the study or maybe the results of the research...in my expectation, the REU program is not very serious...since it is in America, the REU program is very informal.'' This contradicts his mentor who said, ``The sort of relationship that I'm looking for is a professional one.''  However, since they had a different time zone,``We would communicate by email, and then in each email communication I say, if you wanna talk about this, let me know and we can schedule a meeting and that was happening, maybe once every two or three weeks...we were meeting.'' 

%Agreeing to and following group norms could be challenging, even in a typical in-person research experience. Mentees can learn norms by observing mentors and other lab members who usually communicate in the same physical lab space. However, this experience seemed far more different in a remote setting. 

Another example shows how unspoken norms can lead to unproductive student behavior. But when those norms are made explicit, the interaction improves.  Joshua's mentor said, ``I'm trying to push him to learn how to do research...I provided him with a lot of a reference and materials, but after two weeks, he had no single question to ask me...so, I encourage him. Yesterday he started to ask questions. That's very good. So I'm encouraging him to ask a question because if you asked me no question, I assume you know everything.'' Differing cultural backgrounds could lead to differing assumptions about unspoken norms. Both Ivan and Joshua were international students.  These mentees had linguistic and cultural barriers while navigating a US-hosted research program and completing their academic goals. For instance, communication styles can differ within different cultures; interrupting a professor (or any authority figure) to ask questions could be considered impolite in some cultures \cite{zhao2007cultural, rachel2005rethinking, beaver1998adjustment, wan1999learning,hu2002potential, wilkinson2006significance}. This might be more than a language comprehension difficulty and stem from the student's home country's cultural norms. 
%Moreover, mentees talked about different norms in their lab group community, such as, having a welcoming environment, being goal-oriented, collaborative research, and being success-oriented. %
\subparagraph{Academic publishing and goal-oriented group.}
Norms includes things beyond expectations for communication in the group, such as shared values within the group. For instances, some groups were focused on research productivity, including publications emerging from REU students' contributions, while other groups stressed the educational benefits for the students. Frieda who had a research experience in her small home institution said, ``This lab research focused on the publication, which affects the size of the project I am doing.%I could say the centrality of my project to the actual research, so I am doing an offshoot instead of a central thing. Again, that does not bother me because,
...at my school, I would never have worked on something that would be published.'' %During the last interview, she said, ``My result is actually three sentences in their paper. It was a tiny thing which took me a long time to learn because I didn't know how to do it.''

Others groups spent a good amount of time learning about their research background and relevant theories. Grace, who was [simulating] the dynamics of certain molecular systems said, ``There's a lot of stuff I'm learning right now and constantly Google searching. like what is this pigment. What is this protein. Definitely learning. I think I it's really interesting to learn a lot about that and just how intertwined physics is with like supposedly other subjects.''

\subparagraph{Remote work could be more authentic.}
Remote work was more authentic for specific fields. For instance, some fields like high energy and particle physics are primarily done in large geographically separated collaborations and working remotely is part of their norms. For example, Frieda said, ``High energy physics usually work online. They want me to collaborate'' and because of this, ``I definitely consider myself part of the physics community.'' %Helen said her lab group community ``bounced ideas off each other easily, and it seemed their goal was building knowledge, not who's right, who's wrong.''

\subsection{Multiple layers of community.}%Communities developed by mentees and other members of their lab}%developing relationships at remote research lab setting can be hard
\begin{figure}[tbh]
\centering
\includegraphics [trim=153 85 300 
85,clip,width=85mm]{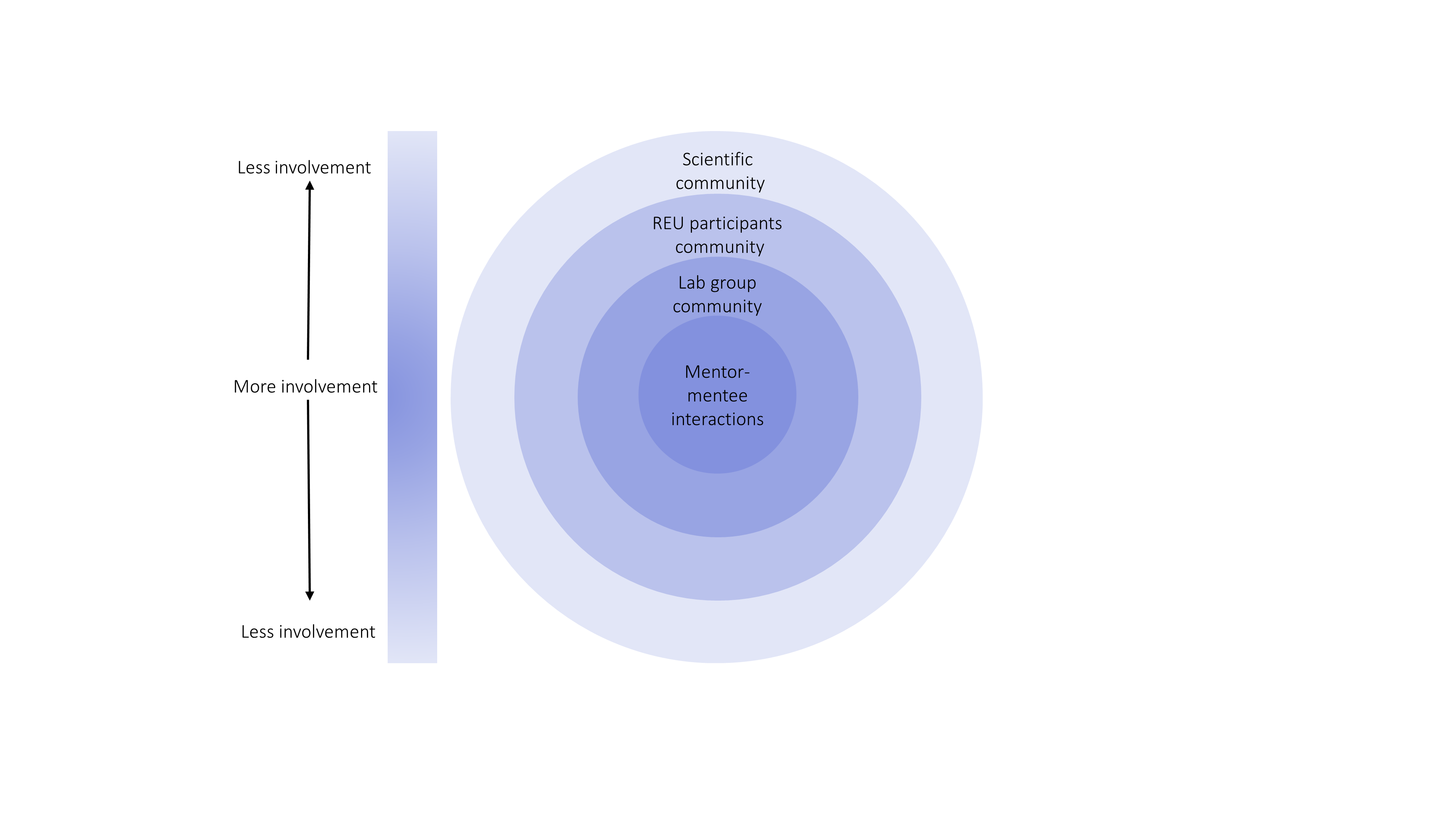}
\caption{The four overall community dimension in the remote REU program. The areas shaded in lighter color denote the lower-level of involvement while the areas shaded in darker color denote the higher-level of involvement.}\label{Fig: Fig_4}
\end{figure}

All in-person UREs benefit from establishing a good relationship between mentees and members of their lab community. Adopting the remote REU program, the need for these relationships is still necessary. However, such relationships could be harder in such a remote space. Similar to any research program, remote REU programs also provide various community levels for mentees to be engaged. Figure~\ref{Fig: Fig_4} shows four different levels of communities that mentees could engage in during the remote research experience. The areas shaded in lighter color denotes less interaction between mentees and members of the specific community. In comparison, the areas shaded in darker colors denote more frequent interaction between them.
%We highlight several of these classifications in the context of our study that subjects (mentees) got involved in while moving toward accomplishing the object (research experience). 

%%%The depth of students’ integration and emotional engagement in both the academic and the social systems of a college or university can have a tremendous effect on their academic achievement (Tinto, 1993, 2002).
\subparagraph{Communication between mentor and mentee developed gradually in the remote context.}
One of the significant components of a good research experience is having a good mentor-mentee relationship \cite{chemers2011role, byars2015culturally, dolan_toward_2009, laursen_undergraduate_2010}. Successful mentoring happens when mentors and mentees are prepared to establish a positive relationship with each other, and the lack of in-person contact made this process slower to develop. David said, ``My mentor and I communicate regularly, and since we have meetings every day as well, we are able to go over things and talk about that very easily.'' Some mentees stated that as they became more familiar with their project and established a good relationship with their mentor, their comfort level increased. %For instance, Caleb said he feels a lot more comfortable now that he has a couple of weeks of experience on it. %`but at first, like you're coming in as a new guy with basically zero knowledge on it and everyone else has like a year or two. I was a little shy but now that I have a little experience. It's easier to kind of voice my opinion.''
When Frieda talked about her mentor, her comments included, ``He explains stuff to other REU students and me. He answers our questions...he gives us resources...he is very responsive and helpful...and we meet almost every day.'' This is a good example that suggests having a supportive, encouraging, and reachable mentor can help mentees establish a beneficial relationship. Five mentees talked about how having a collaborative, independent, and communicative relationship with their mentor facilitated their research progress. Caleb thought his mentor was an excellent example because, ``He just guides you enough that you are on the right track, but then gives you enough freedom to kind of explore and figure it out on your own. But I mean, if you get stuck, he is more than willing to help out and kind of push you in the right direction.'' Several mentees ($N$=6) stated the importance of having a quality relationship with their mentors in addition to how their mentors provide various resources and guidance. Emma mentioned she is very comfortable talking to her mentor and suggesting anything because ``He is very open to suggestions.'' It seems like her mentor was aiming to support the development of the agency. % I pretty much lucked out with the mentor situation. He will listen to anything that I have to say. It seems like he will value anything I have to say. And then if there is issues or reasons, that it does not work, he is very good at not making me feel like I am a stupid little girl, which can sometimes happen with physics professors.'' 
She also emphasized personal aspects that were in common between her and her mentor and how this remote experience contributed to her growth. She said, ``He is a great mentor'' because ``He is socially and politically aware. He does not shy away from discussing politics and stuff with me. He makes me want to be a better person, which I like to be encouraged to do more than I'm doing now. He literally is one of the people who like the Black Lives Matter community...%There's like really bad stuff going on right now. And so they called out to the non black community to come and just like basically stand around they have like  That you're allowed to carry guns openly in in this city. I don't know about the state, but in the city.  And so basically, they just had a call out to anyone who was with the movement and had a gun and could like just make like a standing wall of like not holding guns just showing that they are currently  Like carrying a gun to prevent the other people like who are against the thing. So he was one of the people like standing there with a gun and like  He's not afraid to like  Show. whereas I before may have been really nervous about that kind of situation like he's really encouraged me to   Move outside my shell of thinking, like, the only difference. I can make is by becoming a teacher and like promoting that kind of stuff and education in the classroom.  So yeah, he's definitely  like helped me become even like even more politically active than I was. like we've been to rallies and stuff, but I'm just part of the rally. I've never been like voluntarily like a thing that could really get me in trouble. 
%Whereas I before may have been nervous about that kind of situation like 
He's encouraged me to move outside my shell of thinking...he helped me become even more politically active than I was. When I would meet with my previous mentor, I would sit in his office, and he would tell me what he wanted me to do next, and then I would do it. When I met my current mentor, we talked about what is going on in the country right now. We talk about how are you, how is the puppy. Then, we talked about how [research] goes? What are we going to do next? I think those conversations are helpful.'' %I think it is also important, like maybe not as important for me who's already pretty set in like my values and that kind of stuff. But if I was a young like research student If I was, some of the students who are participating in this REU like 20 years old. I'm 27 so I pretty much know the values but I remember being 20, and at 20 there's all kinds of things that are floating around in your brain and if I had a mentor, like a research mentor. That's telling me like there's this movement going on the world. And it's very important like let me tell you'...
Five mentors talked about having informal but professional, off-topic conversations as a way to promote their relationships. For instance, Frieda's mentor said, ``We met for an hour, every day, almost with the other students and me at some level, we just joked about things, talked about life in general...we had discussions about how things are going on with COVID-19?''
Besides, mentors usually provide more one-to-one academic advice and guidance for mentees' project to their mentees due to a remote setting. For instance, Bruce said, ``I do feel I've been getting a lot of guidance...[my mentor] tells a lot of stories about his past graduate students, and he's talking about what they're doing now...there are just so many different ways; they're either professors working in research labs or working in different things.'' Although, at some point during the program, Bruce expressed concern about the large amount of time that his mentor dedicated to giving him advice about future career paths, by the end of the research program, he ultimately found that information very useful and insightful. He explained that guidance helped him clear his mind around doing research as a future career. This example is one of the many examples from nine mentees that stated how these professional development advice helped them to achieve their goals for these remote REU programs. %Each mentor had their own methods to evaluate mentees' outcomes and provide feedback to them. For example, several mentors spoke to their mentees and asked them different questions about their results and background knowledge in different stages of their project to assess their knowledge. Bruce's mentor said, ``I could immediately see from giving him certain things to read and finding out what he was able to read and understand and implement. I could immediately figure out his background knowledge. That was very clear within a few weeks that what he knew and did not know was very easy to find out from discussions. By giving him certain known problems to trying to verify his codes and his numerical algorithms, one could immediately see what he was able to succeed in and what was working and what was not. So I think those diagnostics were not so hard once he started reading the papers and the parts of the thesis that I sent him and so on. I could really gauge quite quickly where his strengths were what he knew, what he did not know, and what he needed to learn and explore in order to try to work on these research problems. So it was very interesting, but not very difficult to find that.'' Some mentors preferred doing assessments by communication and providing feedback on results in weekly meetings. Andrew's mentor stated, ``There would be plots, and there would be programming, there would be a discussion about the meaning of the plots and things like that. In the language and the quality of the plots, you could see that he understood more and more than the message and how to pass the message and what he learned that was visible.''
%Three mentees stated that they were still establishing a relationship with their mentors. %[olivia, shuijie, anothony]
%\textit{``What kind of relationship. Well, it's supportive, He explains stuff to me and us to Caroline as well. He answers our questions. He gives us resources to us. He is really nice, friendly, whenever we contact him with questions, he like figures it out like he's very responsive and hopeful.  I would say, comparing to maybe other groups, I think that we have much more contact. It sounds like some people meet with their mentor only once or twice a week, but we meet almost every day.  when we leave, we have a plan for what we're going to get done.  Which I like it. That's a structure that's very helpful for me.''}
%Mentors ($N$=2) were very clear about their expectations and their long term plan for the project and they provided mentees a bigger picture of their project.  
%Overall, there was growth in the quality and frequency of the mentor-mentee interaction with passing time. However, the remote settings made this slower to develop.

\subparagraph{Involvement in the remote lab research community.}
\label {Sec: lab community}
In the context of this study, the lab research community refers to members of the same research group. These interactions usually happened once or twice a week, in addition to spontaneous extra meetings based on mentees' needs. Mentees talked about different aspects of their lab community dynamics in the norms section. Because of the positive norms that we discussed earlier in section \ref{SEC: Norms}, most mentees had positive feelings about their remote lab research community, both emotionally and practically. In particular, two mentees spoke about how everyone in their group ``wants them to succeed'' during regularly scheduled Zoom meetings. Grace said, ``They are trying to help each other. It is very collaborative, which I enjoy. It feels a little bit more safe and open. I think it's very success-oriented.'' In such an environment, every conversation can provide an opportunity for mentees to learn and feel they have a chance to contribute to the discussion.

Half of the mentees worked collaboratively with their group members, other REU students, REU students from previous years who worked with the same mentor, or researchers from outside the REU institution. %Mentees stated that while they worked collaboratively together, they were a lot of self-accountability and independence.
For instance, Caleb's mentor said, ``I have colleagues in the field who I collaborate with. I contacted my colleagues, ones in England, ones in Greece, and [Caleb and my previous REU student], we are all meeting today...Caleb will meet world class scientists this way. Today Caleb and my previous REU student talk to these two international well known collaborators, they're going to explain what they've done so far. That's a very good experience.'' The other half of the mentees worked alone on their project except for input from their mentor. Ivan said, ``Another REU student and I work from two different sides and do not have a lot of interactions.''

Similar to in-person REUs, some groups used group meetings as a safe place to provide feedback and swap advice between group members. Mentees ($N$=5) were comfortable sharing ideas with their lab research group in the remote lab format. Grace said, ``We had a conversation about collaboration at the beginning of the REU between us and how beneficial collaboration is and working together. So, I feel pretty comfortable.'' 

Two mentees stated that they felt less comfortable sharing ideas and receiving feedback in their lab group meetings. Caleb admitted that he was shy initially to reach out to other people because other members of the research group had far more experience. Bruce said, ``I am always like in this state, where I am not sure if the goals I set for myself were too ambitious, or did I just not get enough work done. That is kind of like this limbo state always. I think that might be a little bit of imposter syndrome...it might be just a fear of looking stupid.'' Interestingly, in the last interview after the REU program finished, Bruce said, ``I have become more comfortable, and because of that, I would become less stressed about interacting with the community.'' Overall, half of the mentees were very comfortable in terms of seeking help from other lab members remotely. Despite the remote format, mentees were still able to contribute to meaningful research. 

\subparagraph{Lack of REU participants community in the remote context.}
\label{Sec:Lack of REU participants community in the remote context}
The next level of engagement was between mentees and other REU students. When we asked mentees about overall remote communication(e.g., mentor and lab group), many reported it as a reasonably positive dynamic. However, they were also concerned about developing relationships with other REU students in a remote research format. Early in the program, Bruce said there was a lack of communication between other REU participants, ``I have met the other REU participants on Zoom, but it does not seem like there will be much of a conversation happening...like currently, we have a Slack, but no one is commenting on it.'' Frieda had a similar feeling. She said, ``It is going well. Everything is well communicated. I wish that I could be more present with the other REU students because I feel that they are the ones I am not getting to know as well. I could have that if we were in person, and it would be nice to be in person with the professors as well. Nevertheless, I do not think that is as important because we still have constant contact. It is the chance encounters that we are missing out on.'' Helen stated, ``In the beginning, we are all really trying to chat over Group-me to get to know each other and then go away as we got busier [with research]. But I still think that I built good connections and gained a good insight into other students' next steps; where and why each was trying to pursue this REU.''
%However, -----For instance, Ivan said, ``I work on the one side of the project. The whole project is a very big thing. Another REU student and I work from two different sides and do not have a lot of interactions''.
 %For instance, Grace said, ``The REU students get together a lot just to discuss the papers and what we know so far, questions we have for the professor, things we are struggling with, so kind of a mix of both. We are collaborating in the sense that we are all kind of trying to work it out for ourselves and then helping each other out with questions.'' 
Four mentees from different project remarked that working as a part of a team improved their communication skills. Andrew, who felt active in discussing with other REU students in his lab, noted, ``The sort of dynamic we had over just discussed things back and forth throughout the day kind of made me more sociable. I was not worried about getting on a Zoom call with them and talking to them and like discussing things and helping someone if they had like a problem with their code or something. So I think it advanced me.''
%grace, Anthony
%Unfortunately, the lack of REU participants' community persisted throughout the REU program. During the 3$^\textrm{rd}$ interview, Bruce felt ``there is no [students] community.'' According to him, during social events, the REU participants did not interact because they were listening to lectures, and he wished they had some sort of ``all groups' meeting'' where they could present their work to each other.
%emotional engagement has often been conceptualized as students’ sense of belonging, which
\vspace{-0.3cm}
\subparagraph{Few mentees felt connected to the more extensive scientific community.}
Due to COVID-19 and social isolation, mentees had limited opportunities to engage with the broader scientific research community (e.g., other people working on the same topic) or cross-campus activities (e.g., end of summer research symposium). However, we found that a productive and trustful relationship with their mentors and lab group community fostered a sense of belonging in the wider scientific community, even if there were not opportunities for substantial interactions with that wider community. Our earlier framework for psychosocial growth during an REU presented how multiple factors in the remote REU program impacted students' psychosocial gains (e.g., sense of belonging, self-efficacy) \cite{zohrabi_alaee_impact_2022}. This growth takes them to the next level of community which refers to mentees' perception of self within the more extensive physics or research community beyond their lab research group, which may impact their future career outcomes positively. 

Four mentees stated that working as REU student in a group gave them more confidence and a greater sense of belonging to the scientific community. Helen, reflecting on the possibility of being in a larger lab in graduate school, said, ``I can definitely feel more confident because I have something to compare to or talk about and reference as an experience. I know what it would look like if I was to pursue that.'' She continued, ``It is easy to have research in an undergraduate institution because no graduate students are competing for these positions. But it also means you are usually not part of a larger research project. In terms of that, it was awesome to see what that looks like and what it looks like to be part of a research group and like a larger lab setting where you have to sign up for lab time and plan your experiments in very detail because you only have a certain amount of time the lab. That is something I do not really have a problem with at the undergraduate lab. My lab is my lab. So I think it was insightful to see how that dynamic operated. I still, as I said, gained good network connections, even though it was remote. We did not have social interactions every day; it was still an opportunity to meet new people and capitalize on those connections.'' %Mentees' emotional state development as a result of the activity system would happen through social interactions with other members of the community and impact mentees' psychosocial outcomes \cite{zohrabi_alaee_impact_2022}.

%Yusen Feng: I think it doesn't matter.  But if you talk about the concerns I really worry about if I didn't see my classmates in the REU program. And maybe

%Shuaijie: Actually, I am very want to go to Indiana, but this will. This is kind of canceled, so we can only talk through the internet. My concerns. And my concern is maybe, maybe the internet is not so steady and. And it's hard to imagine that, talking, talking with your face to face, or talking with my professors or dr tan or other people face to face is a better way to communicate with each other.
%Experiences that take place during UREs, such as->nrc

\subsection{Division of labor among mentors and mentees}
The division of labor construct refers to how set of tasks and responsibilities are shared among members of the community to achieve a common goal. For instance, the mentors' role was to provide mentees with resources and feedback, while mentees were responsible for completing their projects, sharing results, and receiving feedback. However, COVID-19 changes many aspects of the division of labor among both mentors and mentees. In an ideal in-person REU program, the labor is more likely divided among REU students and graduate students and postdocs who are working with the same mentor on the same project for the same goal. Interestingly, still, in a remote setting, half of the mentees worked collaboratively. They stated that while they worked collaboratively together, they were a lot of self-accountability on their part. Grace explained that their common objective is to ``learn''; she said, ``The REU students get together a lot just to discuss the papers and what we know so far, questions we have for the professor, and things we are struggling with, so kind of a mix of both. We are collaborating in the sense that we are all kind of trying to work it out for ourselves and then helping each other out with questions.''

Similarly for Frieda the division of labor mediates the community's relationship to the object which is learning in order to process some new results for the project. She described that ``When I don't have information and I don't know how to get it. I usually email my mentor or the graduate student, depending on what I need information on...I currently have to ask for help a lot which is kind of expected. They all expect me to not fully know what I'm doing, because  I've never done this before. And they're teaching me, which is the point of an REU, but it's slower than I would like. And I have to rely on people more than I would prefer. But it is, I'm learning a lot. It just takes a little longer...the [graduate student] sent me a lot of information last night, which is a documentation, because there's no documentation for this code...he made a specific analysis in [specific program to produce physics plots]...I will probably know by the time I'm done reading that what I need to know and if I don't, then I'll email him again.''

As we mentioned earlier in section \ref{Sec: lab community}, half of the mentees worked alone on their project. As a result of the lack of face-to-face interaction during remote research experience, it is hard to organize the community members to meet common goals, and mentees only consider their project's finite objectives and their own role in the REU program. This attitudes is followed by procedural steps and independent, self-paced learning. 
The sudden change in the format of the REU caused mentors to be burdened with research design, and they had to modify a research project with no laboratory work and different forms of social interactions. Due to the nature of these new remote research settings, the project design eliminates the need for most division of labor. This is why six mentees stated they felt independent in doing research since they performed different roles and had less chance to share knowledge with their group members, however this could also be a sign of increasing ownership and self-efficacy among students. %Helen said, ``I feel more confident since I have worked so independently, and it is definitely nice to have a mentor to direct you and like what you are doing to make sure that the data is a good representation of what you are saying it is and then making sure it is beneficial. So I think I am confident, but then obviously there is some mentor person to help guide you.''

%alone: Olivia-Anthony-shuaji-yusen-mckeena

%Six mentees were often comfortable in sharing ideas in the remote lab format. Grace said ``We had a conversation about collaboration and the beginning of the REU between all of us and how beneficial collaboration is and working together. So I feel pretty comfortable.'' Two mentees stated that they feel less comfortable sharing ideas and receiving feedback in their meetings. It was different for Bruce at the beginning of the REU program; he was reluctant to ask questions and share his results with his lab group. He admitted that he had a ``fear of looking stupid.'' By the end of the the REU program, he said,``I've become more comfortable and... I was also able to ask questions online, essentially, so I'll be able to ask my mentors for help when I got stuck on something. They were able to help me out a lot. They're just able to give advice that worked really well.'' On the other hand, Bruce's mentor said, he tried to ``reply back to him quickly.'' Likewise, Helen, felt more cautious in sharing her thoughts and ideas with her group. She said, ``Sometimes over zoom it can be easier because you're not actually in person, so it's easier to just say what you want, but then also it's harder because you don't know people as well.'' 
%They refer to themselves as an independent researcher
%indep->ownership

\section{Research question 2}
\vspace{-0.5 cm}
\textit{\textbf{What challenges we observed within the goal-directed REU activity?}}\\
\label{Sec: Challenges rq2}
COVID-19 impacted both mentors and mentees who participated in the remote REU program. They all joined this program with their self-expectations, concerns, and emotions about the new remote research experience.
%The most common concern for both the mentor and the mentee was working remotely. %Both groups said they would miss the lab community. Caleb said his biggest concern with a remote experience was the ease of access to everything. He said, ``We don't have a lab we can go into. You have to set up Zoom meetings to meet with the adviser instead of just going down the hall to his office. A lot of stuff, in my opinion, is easier in-person. But with Zoom and the technology we have today, it's about as easy as it can be to do things virtually.'' In addition to remote aspects of the REU program, we noticed that two female mentees and one male mentee talked about imposter syndrome in the first weeks of the program, which could be the same concern in both remote and in-person formats. Grace said, ``A little self-doubt every once in a while like how did I land this, am I really good enough, am I going to know the material well enough? Do I know enough physics beforehand to do well in this position?''
%As we discuss challenges, we acknowledge that COVID-19 pandemic forced rapid changes to turn these experiences into remote opportunities. 
Some challenges may be particular to the remote format, while others may occur in any research opportunity. For instance, Grace said, ``It went pretty smoothly for being a remote internship. I know we were very delayed and getting our access to the supercomputers that are beginning, which was a little frustrating, but that is kind of the only thing I can think of.'' This challenge was not only due to the remote REU format because the supercomputer access is typically remote. We give several examples of ways in which the remote format caused extra challenges for mentees.

\begin{figure*}[t]
\centering
\includegraphics [trim=160 175 25
100,clip,width=\textwidth]{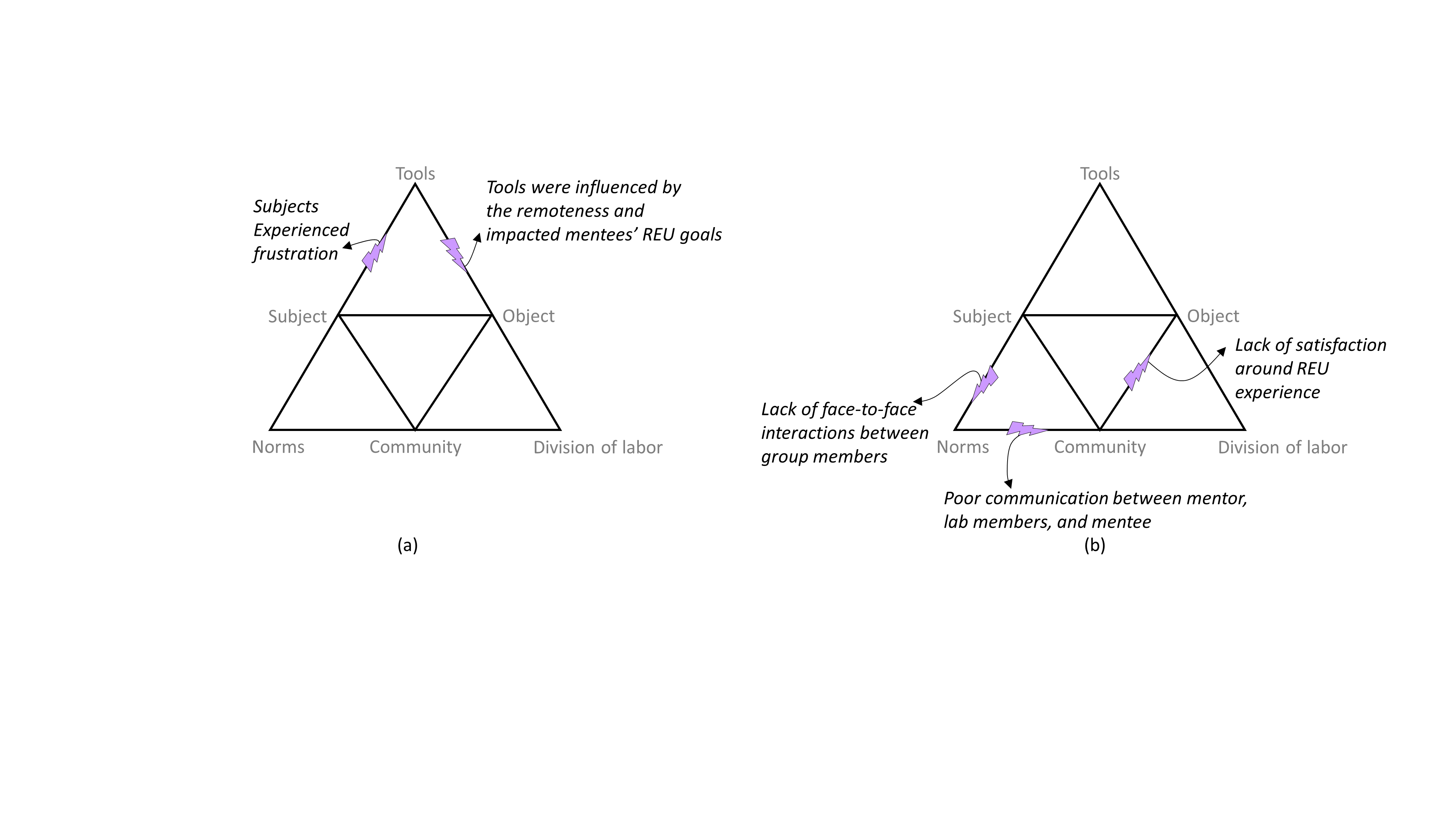}
\caption{The ``tension'' icon denotes the impact of contradictions between two linked components of the activity system that could affect the outcomes of the REU activity.}\label{Fig: Fig_5}
\end{figure*}

\subparagraph{Facing technical challenges at home.}
Some of the common challenges mentees faced with remote research format were Internet connection and technical difficulties. For instance, Helen had some difficulty to ``compile different software. That was not really my fault. It was just my computers on my setup solution. So It was just mostly a miscommunication through a remote experience...I have been frustrated with the whole software thing, but it is also like fulfilling when you finally figure it out.'' This example explains the norm and expectations around the importance of working through difficulties and identifying the challenge of working from home and limited access to seek help. Figure~\ref{Fig: Fig_5}(a) shows the tension that arose mainly from limitations of accessing tools and achieving their goals. Again, from the CHAT perspective, there was a challenge around mentees employ tools in their research process to participate in the REU activity and achieve their goals.

\subparagraph{Working from home introduced motivational challenges.}
%you are at home no physical interaction
As we mentioned earlier in Section \ref{Sec: Constructing the lab environment from home}, the sudden transition from working in a lab environment to working from home has introduced some challenges. Helen claimed that doing research in the remote setting was hard, ``Because different things happen in your house or the world. You just don't have that same motivation sometimes from your environment...some days it was harder to do as much work.'' Bruce said, ``Be[ing] personally motivated was kind of difficult [when you are working from home]...because I was living at home. There were certainly some things that were distractions.''
%We also noticed, another female mentees and one male mentee talked about imposter syndrome, which could be the same concerns in both remote and in-person format. Grace said, ``A little self-doubt every once in a while, like how did I land this, am I really good enough, am I going to know the material well enough? Do I know enough physics beforehand to do well in this position?'' 
%Bruce and Helen cited motivation as an issue. "There was certainly some things that were distractions on the first thing was just be personally motivated was kind of difficult.
%Activity systems are characterized by their internal contradictions (Engeström, 1987, 1993;Leont’ev, 1974). These contradictions are best understood as tensions among the components of the activity system.
%These tensions or disturbances do not occur accidentally or arbitrarily but instead are inherent in human activities (Engestrm, 1996). 
%https://link.springer.com/content/pdf/10.1007/978-1-4419-6321-5.pdf
%Researchers find tensions in activity systems when elements from one or more components pull the participants away from fulfilling the activity purpose.These tensions can encourage the activity to collapse or become the reason for change (Engestr?m, 1993)
%Engeström 2001 use 2 activity and highlight the tension of not having a same object between the two subject. this lack of alignment will lead to tension and contradictions that may occure within an interactivity system
%Bruce said, ``Because I was living at home. There were certainly some things that were distractions.''%... Now that I am looking back, I wish I had gone outside more because I really got clumped inside, and that probably was not great for me. I felt like my mind got stuck sometimes because of that.''   
% tension from family 
Bruce also talked about the tension between him and his parents regarding the possibility of taking a gap year before starting graduate school. He added, ``They also wanted me to take the gap between my years at college. They wanted me to take a year off between college to do research or do something.'' Figure~\ref{Fig: Fig_5}(a) as an example of diagrammatic representation, shows that this challenge around the physical research lab environment and challenges with living with his family can impact Bruce as a subject of the REU activity system as well as his process towards achieving his goals which were ``to figure out if I wanted to spend time in a research setting'' and applying to graduate school.

\subparagraph{Information overload and lack of time.}
Another challenge was the lack of time and learning lots of background material. As one example, Joshua had experienced frustration; He said, ``The reason is that I need to learn a lot of basic background knowledge. So, that is hard.'' Bruce said, ``I am pretty limited on time...I know there's a lot more information to go on it, but I just haven't really done that and dissuaded from moving in that direction, just because of the limited time I have.'' 
%\subparagraph{Doing summer classes and undergraduate research at the same time.}
%For Joshua, one emerging challenge was managing his time between summer courses in his home institution and the REU program goals. He said, ``I have to take two university courses at the same time. About two weeks ago, I have to go to another comparison university to finish it, it's a report or what we call a lecture to other people. Before sending my lecture and my report, I have to fix just one bug in my program and I have to fix it before I go to bed. [At the same time] there [was] a workshop in the REU program, I have to cancel it and try harder to fix that bug. It was 2 o'clock in the morning while I was still in front of my desk and working, and the professor sent me an email and told me that ``you have to attend to those [workshops, they are important [...] and if you're just absent again we have to get you degraded.'' That was really awful. I really have to make a choice. If I am going to attend to those REU program's [workshops] or debug my code bugs [...] it's a really hard time for me to balance to universities things.''

\subparagraph{Barriers can arise in mentor-mentee relationships.}
Analysis of the data identified that most mentees maintained a positive professional relationship with their mentor. However, two mentees talked about two barriers. One barrier noted was when the mentor was unavailable to provide support, guidance, and feedback. For instance, Helen said, ``If I was not hearing back from my mentor was hard to know what to do. Then you kind of feel like I'm not really putting as much effort as I should be because I'm getting paid for this and should be doing this. So just that mental aspect of you are not doing enough on like some down days.'' 

Barriers also arose when the expectations of mentees appeared to be different from mentors' expectations. Ivan claimed his research group was not serious and formal. He said his lab members were ``Very kind and funny, [but] it is just different from my expectations.'' He wanted a chance to take ``some courses'' and wished his mentor asked him to bring ``some ideas about doing some research.'' While his mentor thought ``He might have been a little bit disappointed with the research project he had because I think he wanted to look at big overarching questions, and we wanted him to look at the data that came out and some checks...we all have dreams of sitting in a room and dreaming up some theory that explains everything, and there is nothing tedious about that process. However, when you actually get to work on an experiment, that is tedious, and it is required to make progress. You have to learn a certain amount of discipline and commitment to seeing through the tedious part so you can get to the fun parts of having a discovery or learning something.'' We would like to acknowledge this is a more general issue that could happen in any form of REU.  The development of a mentor-mentee relationship in a remote setting takes time and effective communication between mentors and mentees \cite{Zohrabi_perc2021}. Ivan was a Chinese student who participated in the REU program. It was not easy for both Ivan and his mentor to schedule meetings due to different time zone. His mentor said, ``I could have sat him down every day, and we could have had a daily meeting where I demanded that he focus on his research and show me what he was doing, but that was just not feasible in my schedule to do that.'' In addition, his mentor did not feel comfortable grouping him with other REU students since they had different levels of knowledge and worked in different time zones. Figure~\ref{Fig: Fig_5}(b) shows that the lack of communication between mentors and mentees can cause tension in their relationship, which impacts other aspects of their experience. Mentors and mentees can develop better community support by discussing their expectations, group norms, and rules.
%so we have regular meetings... We would communicate by email, and then in each email communication and say, If you wanna talk about this, let me know and we can schedule a meeting and that was happening, maybe once every two or three weeks for him. For the other students, we were meeting at least once a week and sometimes multiple times a week... It was just really hard for me to justify putting in more time with him, given the amount of time that I seem to be getting from him. I mean, I could have sat him down every day and we could have had, a daily meeting where I demanded that he focused on his research and show me what he was doing, but that was just not feasible in my schedule to do that. And at the end of the day, if a student is not motivated to to commit a certain amount of time, I don't think it's helpful for me to try to basically compel that sort of motivation... 

% between reu students and coordinators- other group
\subparagraph{Poor communication among REU students.}
\label{Sec: Challenges-Lack of REU community}
As we mentioned earlier in section \ref{Sec:Lack of REU participants community in the remote context}, knowing that a lack of REU participants' community persisted throughout the REU program might make REU coordinators and mentors develop new ways to creates virtual communities of REU students for the next time. During the 3$^\textrm{rd}$ interview, Bruce felt ``there is no [student] community.'' According to him, during social events, the REU participants did not interact because they were listening to lectures, and he wished they had some sort of ``all groups' meeting'' where they could present their work to each other. Although most of the mentees had a weekly interactive remote meeting, the outcomes seemed far from satisfying. 
Bruce added, ``From the administrative side, there weren't many updates, and there weren't that many check-ins with the administrators with the students. So, there was definitely a pretty large disconnect there. Also, there was definitely no connection between any of the students either, neither students nor researchers. There was no connection there. I think those were the majority of the challenges for me...since the first week, there has not been a single chat in the group chat. I probably remember most of the other people, and I probably remember all their names, but that's the extent to what I know about them.''

%tshirt On the other hand, Emma talked about small missing aspects of the remote program. For instance, She stated she really wanted the REU T-shirts or commemorative mug from this experience and finally his mentor gave her couple T-shirts from some art to science activities that he did in the past. 
\subsection{Outcomes}
\label{Sec: Outcomes rq3}
\section{Results on research question 3} 
\textbf{\textit{What are some of the short term outcomes of the remote REU program?}}
REU programs are multifaceted learning activities that involve subjects (mentees) working towards achieving objectives (REU goals) and outcomes through mediated components (e.g., tools). 

%REU programs are multifaceted learning activities that help students pursue careers in STEM. CHAT framework provides a broad insight to the components of these research experience that may influence students' outcomes such as persistence in STEM.% For instance, how different communities used by mentees mediate achieving their goals through cultural norms.  
Outcomes are the final part of the goal-oriented REU activity system. 
A few weeks after the program finished, in the 9$^\textrm{th}$ interview, we asked all mentees to describe the outcomes of their REU experience. We found that almost every mentee, despite being from different demographics with different projects, achieved their goals as a result of their positive research experience. They all stated that attending the REU program was the right decision in their academic journey. The following parts explain some of the specific short-term outcomes of the remote research experience.

%\subsubsection{Self-reported growth in mentee's retention in STEM fields}
\subparagraph{Mentees were able to clarify their future career decision-making.}
%This result is consistent with the evidence in substantial prior research that has examined the influence of UREs on students' outcomes, such as clarifying future career goals \cite{lopatto_survey_2004, seymour_establishing_2004, hunter_becoming_2007}. 
%Many mentees talked about how participating in this research program helped them figure out their interests and narrow down their future career options, including graduate school. For instance, Ivan said he is interested in studying nuclear physics in graduate school, but ``before the REU program, I did not even know what I was interested in because there were too many fields.'' Bruce said, ``I learned something very helpful. I gained research experience. I feel like it gave me a little more focus, and it has given me a choice. I have decided that I want to pursue a Ph.D. and try to research a lot longer.'' During the last interview, David, who plans to apply for graduate school, said, ``I am just able to experience what research in physics is and get a feel for how much I enjoyed that.''
The remote REU experience positively impacted most mentees' future career decision-making. Most of the mentees ($N$=8) said the REU experience helped them understand the nature of research work and think about their future career in STEM field. David stated, ``Especially because I was able to get a better idea. Not only I was able to learn more about the software and about the research going on in one of the fields that I am interested in, but also because I was able just to get a better understanding of how research works and that sort of dynamic. I get a better feel for what doing research with a professor looks like. So that I have a better understanding of how that will work when I am, for example, in graduate school doing research.'' As part of Caleb's REU experience, he learned that he could not be on a computer for eight hours a day. He explained he enjoys computational theoretical research, but ``I would like to be in a lab-type setting; some days, I am doing hands-on research, and other days, I am doing computational stuff. My biggest takeaway [from the REU program] is that I do not think I could just be on a computer for eight hours a day.'' Due to the lack of information about different career choices and fields, some mentees were unsure about their future job. For instance, Ivan said he is interested in studying nuclear physics in graduate school, but ``before the REU program, I did not even know what I was interested in because there were too many fields.''

\subparagraph{Mentee is going to apply for graduate school.}
After finishing the REU program, most mentees were sure that they would like to apply to graduate programs. For instance, Frieda said, ``It was really good for where I was in my education and made me seriously consider this field as a graduate school field, which not that I was not considering it, but I am much more serious about considering it now.'' Bruce, who was unsure about his future career during most weeks of the REU program, said that his mentor provided him with a new perspective and insight on doing research. He said, ``I definitely think that [my REU mentor] has a strong influence on making my decision about graduate school.'' Bruce explained that this influence came from stories that his mentor told him about his previous students' career decisions and also from giving him a sense of how he does research. He said, ``There's just so many different ways that they're either the professors or they're working in research labs or in different things...it's just stuff that I really would like to be doing in the future as a career...I think another [influence] is just he's giving me a sense of how he does research...he's not always super stressed about everything. He's not overworking himself. This idea that you can be a researcher and not overwork yourself all the time is also quite enticing.''

\subparagraph{Mentee engages in new learning opportunities in the STEM field.}
Some mentees explained how to tailor their interests toward their future career goals by choosing elective classes, attending the seminars, or reading new related articles after the REU finished. One example was Joshua, who decided to take some elective courses in astrophysics and nuclear physics the following semester to narrow down his interests. Other mentees tried to read and search through the literature related to their summer research experience. For instance, Helen said, ``I like to read articles. I am part of the American Nuclear Society and get updates and stay informed, especially around medical physics. I always have liked reading Physics World news about medical physics, listening to podcasts, and connecting with people. So I have definitely been more interested in learning more in that area after that REU.'' Similarly, Andrew, who did some research around nuclear physics during the summer, said, ``[What] I did over the summer made me want to learn more about that...I got a book on nuclear physics, and I was kind of reading that as the program went along.''

%\subsection{Promote STEM disciplinary knowledge and practices}
\subparagraph{Mentees learned physics content.}
This remote research experience deepened mentees' knowledge in discipline-specific project content and also introduced them to new areas. For instance, Grace said, ``I am enjoying it a lot. I am learning a lot about chemistry and biology and just how much physics intersects with everything, and how it is fundamental science. I definitely had some frustrations in some of the earlier weeks reading some of the papers and not understanding them...feeling like I had to understand everything that was being thrown at me. Overall, I am definitely enjoying learning these, learning how to use the program, learning  the chemistry and physics involved has been really interesting and enlightening.'' Joshua, an international student, talked about how much he learned from extracurricular activities during the REU program; ``I learned the skills such as organizing the introduction or other parts of a paper better. In GRE preparation courses, I learned some required physics concepts in English...these are important for my future career. In seminars, I  can broaden my horizons, such as I learned some new physics knowledge in other physical areas.'' 

Getting more content knowledge can impact mentees' self-efficacy. For instance, Helen stated that through remote research during the COVID-19 pandemic, she spent ``more time reading different papers about the same thing. I just feel more confident and know it better...I think I have always enjoyed research, and I enjoy the process. I think undergraduate research is important because I have been exposed to many different areas of physics.''
%Besides, five mentors said mentees learn how to use scientific tools during this research experience. 

\subparagraph{Mentees appreciated value of doing research independently.}
Due to the lack of in-person interaction between mentees and other lab members of the community, they experienced lots of independent learning. For instance, Caleb said, ``It makes [the research] a lot more independent. Because you are still able to reach out via Email, Zoom, or whatever platform you use, but there is that added step of composing the email asking for Zoom chat. Whereas in in-person, just like walked out to the office or they are in the room with you, and just like look over your shoulder. So, it adds a lot more independence.''
Likewise, David explained that he became more independent in his learning since, ``I will be in the middle of a project, trying to figure something out in the middle of the day. Then I will just go look it up and try to figure it out. I usually use it as a resource on my own, and I think that might be a bit different if I was working with the professor, more closely in a physical environment.''  

\subparagraph{Most mentees were able to utilize disciplinary research practices.} 
Regardless of the project area, every REU program aims to provide opportunities for students to refine their knowledge, develop research practices, and become community members by supporting them to work collaboratively, solve problems, and think critically. During interview nine after the REU program finished, eight mentees felt more comfortable with their ability to getting involve in their community (e.g., by asking questions and learning new concepts). Mentees join the REU with different academic backgrounds and personal recognition. Three mentees stated that because of friendly and supportive lab group dynamics, they became more comfortable with the level of their knowledge. For instance, Grace said, ``%A little bit more confident in my capabilities and just to accomplish something,
%Ask question: amanda, kate , adam W9-Q14
%mckenna  yusen, bryce brian and anthony feel more comf in the community
[ I can] ask a question and not feel like I am asking a stupid question. It is very hard to look back on my near past self and reflect because...it is hard to see change.'' David said, ``I was just trying to learn to catch up and to understand what was going on...I felt a lot more confident in what I've been able to learn and I had a better understanding of what was going on. So I was able to contribute a lot more.''

%Andrew said as a result of supportive group, he felt``more sociable'' and ``keeping active in discussing''. , I guess like  I wasn't like worried about like getting on a zoom call them like talking to them and like disgusting things and helping someone if they had like a problem with like their code or something.  So I think it advanced me. I like that part. But other than that.  I can't really think of like a big example.
%Similarly, Andrew felt a greater learning experiences encountered throughout the REU program, ``I was able to ask questions online, essentially, so I'll be able to ask my advisors for help.'' 
Six out of eight mentors were very happy and positive with their mentees' progress.

\subparagraph{Remote REU experience increased mentees' sense of ownership of the project.}
In our extensive study, mentees talked about a sense of ownership as a factor that linked to their sense of belonging \cite{zohrabi_alaee_impact_2022} through scientific contribution. In order to achieve the ownership construct, mentees need to produce potentially new results for their project. For example, David said, ``I am contributing to the group and understand what is going on across the group.'' Besides, five out of eight mentors explained that their mentoring philosophy includes helping mentees increase their sense of ownership. Three mentors mentioned giving their mentees the freedom to challenge themselves intellectually. Caleb's mentor said, ``By respecting and giving them great freedom, that is, say if he had ideas on how to do things, I would say. Do it and show me. I did not have to tell them this is how you do it step by step in the beginning. Yes, but it did not take long to get going on his own and then come up with his own improvements and extensions and search. He was very good at that. So I certainly think that encouraged him. I think he liked that. I think he liked the research. But, you would have to ask him, I guess.'' 

Other mentors ($N$=2) provided mentees cultivated ownership by providing the bigger picture of the project and their expectations. As Frieda's mentor explained, ``She had a project, and it was very well specified that this was her project. There was no one else working on it, and again, I am working with graduate students and postdocs, making sure they were not working on the project. So, it was hers and hers alone.'' 

\subparagraph{Most mentees became familiar with the physics community and expressed a stronger sense of belonging, self-efficacy, and physics identity as a result of the remote REU experience.} 
As part of the outcome of participating in the remote REU experiences, many mentees reported growth in their psychosocial skills, such as a sense of belonging, self-efficacy, and physics identity \cite{zohrabi_alaee_impact_2022}. Regardless of the lack of REU participants' community in the remote context, having a good lab research community and mentor-mentee relationship still helped most mentees to exhibit psychosocial growth. During interview nine, almost all mentees reported a higher level of sense of belonging at the end of the REU program. David felt he had a ``place in the field now.'' He thought his research experience over the summer positively helped him feel like a part of the physics community. He said, ``Just because I have been able to experience and contribute as well as being able to look through the work that other people have done in the field a lot more in-depth and having done my little bit of work in the field helps me to understand better the work that others have done in it...even now that I am not doing that research full time, I feel much more in part of that than when I was only taking classes before.'' Both Helen and Bruce, who had a lower level of sense of belonging during the program, believed they felt more belonging after the program finished since they were back on campus and communicating with more people. Frieda said, ``I definitely consider myself part of the physics community...a lot more than I did at the beginning. Because now I feel I know a lot about it, I have gotten a lot done, I am informed, and I feel more in it, like I am more submerged in it personally. Maybe you could say my physics identity has gone from like an amateur to like a beginner to intermediate now. I feel like I am actually part of the field.'' A higher level of sense of belonging and identity is associated with improved academic performance and possibly enhanced persistence in the field of physics.

Findings from analyzing the self-efficacy construct \cite{zohrabi_alaee_impact_2022} indicate that self-efficacy stemmed from various sources, such as getting more physics content knowledge, doing independent research, and producing new results and scientific communication with other members of the community). For instance, David said, ``I was just trying to learn to catch up and understand what was going on. After I had been doing it for about a month, I felt a lot more confident in what I had been able to learn, and I had a better understanding of what was going on. So I was able to contribute a lot more...I feel a lot more comfortable than before because I had a chance to work full-time doing physics, even if it was just for summer. Since I enjoyed that, I feel a lot more confident proceeding with that as a career goal.'' His confidence stemmed from getting more physics knowledge and understanding of his research group projects.

Physics identity is focused on a mentee's perception of self within the physics areas. For most mentees, developing a more robust physics and researcher identity followed after a stronger sense of belonging and stronger self-efficacy \cite{zohrabi_alaee_impact_2022}. During the REU program, six mentees
%Emma, Helen, and Frieda, Ivan, joshua bruce
Six mentees believed they were a physics apprentice who was on the right path and needed more knowledge and experience to be a physicist. After the REU program ended, five
%W9-Q2yusen, anthony olivia, shuaji and kate
mentees felt a stronger identity around physics and doing research in the field. For instance, Joshua said, ``The REU has taught me something about confidence in doing research. The REU has made you think yourself as a physicist.'' Bruce said, ``My focus has changed. I am a lot more focused now, and I am more driven to do the research. And because of that, I am actually looking at different graduate schools now in the process of applying to them.'' Participating in research positively impacted his psychosocial outcomes associated with his future career decision-making.
%\section{Results on research question 4} 
%\vspace{-0.5cm}
%\textbf{\textit{How do outcomes of the remote and in-person research experience compare to each other?}}\\
\section{Discussion}
\begin{table*}[t]
\begin{tabular}{p{6cm}|p{11cm}} 
\hline 
\textbf{\makecell[l]{Goals expressed in literature for\\ students participating in in-person\\ UREs \cite{national_academies_of_sciences_undergraduate_2017}}} & \textbf{\makecell[l]{Short term outcomes of the remote \\Research Experiences for Undergraduates\\based on our data}}\\
 \hline\hline
{Increase mentee's retention in STEM fields}& {-Clarifying future career interest}\\
    {}& {-Applying for graduate school}\\
     {}& {-Engaging to new learning opportunities in the STEM field}\\

\hline
{Promote STEM disciplinary knowledge}
& {-Learning physics content}\\
    {}& {-Appreciating value of doing research independently}\\
      {}& {-Utilizing disciplinary research practices by asking questions and directing projects}\\
\hline
{Integrate students into STEM culture}
& {-Finding interest in physics field}\\
  {}& {-Expressing higher level of sense of ownership of their project during REU experience}\\
   {}& {-Becoming enculturated in the physics community and expressing a stronger sense of belonging, self-efficacy, and physics identity as a result of the remote REU experience} \cite{zohrabi_alaee_impact_2022}\\
   \hline
\end{tabular}
\caption{Comparison between the primary goals for in-person UREs and the actual outcomes for the remote REU program.}
\label{tab: comparison}
\end{table*}

Prior to early 2020, the vast majority of articles about online education debated whether academia should use remote learning or the effectiveness of these methods. In a perfect world, maybe! However, when the COVID-19 pandemic broke out in early 2020, many institutions rapidly experienced a remote learning transition. %Some institutions shifted their summer REU program from an in-person format to a remote format, and both mentors and mentees started working from their homes. 
In this article, we used the CHAT framework to describe different components and outcomes of this new remote complex educational system. Table~\ref{tab: comparison} shows the primary goals for any UREs adopted from \cite{national_academies_of_sciences_undergraduate_2017} (right column) and outcomes of the remote research experience from data we collected (left column). 
%As we have identified in previous section~\ref{Sec: Outcomes rq3} the short term outcomes of the remote research experience are very similar to the goals of many in-person undergraduate research programs. %Table~\ref{tab: comparison} summarizes a simplified primary goals for any UREs adopted from \cite{national_academies_of_sciences_undergraduate_2017} and outcomes of the remote research experience. 
% The short term outcomes of the in-person remote research experiences are usually focus on learning very similar to the goals of many in-person undergraduate research programs.
In addition to the short-term goals of many URE programs (e.g., improving students' learning gains), the overarching goal of many URE programs is to provide a long-term benefit. Most studies have focused on three major outcomes for UREs: retention and persistence in STEM, promoting STEM disciplinary knowledge and practices, and integrating students into STEM culture \cite{nagda_undergraduate_1998, rodenbusch_early_2016, graham_increasing_2013}. In our data, mentees expressed similar long-term goals around their future career decisions. Also, they described the short-term outcomes, such as preparing graduate school application materials and meaningfully engaging with their lab community. In the following section, we review the three main URE goals and discuss them individually in light of our data. 

\subparagraph{Increasing mentees' retention in STEM fields.}
The UREs can impact students' retention in STEM majors and provide them with a new way of thinking about their future career paths in STEM fields. In-person UREs often aim to help mentees understand what it means to do research and what a science career might look like \cite{national_academies_of_sciences_undergraduate_2017}. The outcomes associated with this goal across our data include: clarifying future career interests; applying for graduate school; and engaging to new learning opportunities in the STEM field. This result also is consistent with the evidence in substantial prior research that has examined the influence of UREs on students' long term outcomes, such as clarifying future career goals \cite{lopatto_survey_2004, seymour_establishing_2004, hunter_becoming_2007}.

For instance, The comments Andrew made during the interviews provides insight on some of the factors that may have contributed to his retention in STEM fields. He said, the REU program ``has influenced me in the fact that it makes me want to do graduate school and research more. Because beforehand, I was like, ``I do not know if that is going to be a lifestyle I want to get into, because I did not know. I have not done research before, so I did not know if it [would] be something I enjoyed or something I [would] absolutely hate. So after doing this REU, I was kind of like, I enjoyed this. I can see myself doing that.''

\subparagraph{Promote STEM disciplinary knowledge and practices.}
The UREs can help students to develop new skills, learn new knowledge, and engages in the practices of their STEM discipline, such as using computational models or analyzing and interpreting data. The outcomes associated with this goal in our data include learning content knowledge through the REU experience, gaining independence as a researcher, and developing questioning skills. In terms of the activity system CHAT triangle, the disciplinary knowledge and practices would be considered research \textit{tools} that support the research objectives of a student's project.
Several previous studies on UREs outcomes indicated that meaningful research practices could facilitate students' learning and development of their technical knowledge in addition to promoting their communication skills \cite{junge_promoting_2010, hunter_becoming_2007, kardash_evaluation_2000, national_academies_of_sciences_undergraduate_2017}. Similarly, in our data multiple times during the interview, mentees said they felt like they better understood the physics concepts and what was going on in their lab research group. Besides, most mentees were very comfortable reaching out to lab members (including other REU students, graduate for help. %Frieda said they don't talk me down

\subparagraph{Integration students into STEM culture.}
Other studies of in-person UREs mentioned that undergraduate research experience could strengthen the students' motivations and interest in STEM culture. Here, integration into STEM can be thought of as a psychosocial growth process (e.g., a gain in physics identity and sense of belonging) where  students learn practices, tools, and values of the discipline. Our findings around integrating mentees into STEM culture include a developing interest in the STEM field, a sense of ownership of their project, and growth of psychosocial constructs. There is considerable evidence across our data that positive remote REU experiences require significant community interactions to achieve desired outcomes. The base of the CHAT triangle (i.e., norms, community, division of labor) was referred to as the social basis of the activity system \cite{engestrom_learning_1987, engestrom_expansive_1999, engestrom_learning_2014}, which was more challenging to form in the absence of in-person social interactions. However, our study shows that these social bases can take different shapes and qualities in a remote situation (e.g., using the Slack app to communicate with other community members). %For instance, Andrew said, ``I tried to contact [other REU students] multiple times a week because obviously, they are online too.'' or  Freida's mentor thought the remote aspect of the REU was not a big challenge. He said, ``If she were here, she would be sitting in probably the available room with REU students, she would be sitting in her room on her computer working eight hours a day... And If she was here, she had exactly the same project she had. The only difference would be informal small office interactions.''
Nevertheless, the lack of community and communication with other REU participants persisted throughout the REU program. One way of enculturing students in the community is through strengthening interactions between mentors and mentees. Emma, who had previous experience working with an in-person mentor, said, ``I really enjoyed working with my [REU] mentor; I think the dynamic is great. He was my advisor, but he also treated me like an equal, which is uncommon when working with anyone with a doctorate when you are an undergraduate student... My [REU] mentor trusted me to get things done and to have my own ideas. He kind of believed in my ability to do any physics-related thing, which is not super common when people see that I am instructional physics... My [REU] mentor was just like, here is the stuff we are doing, what do you want to do... And he was not like, do you know how to do circuits? Do you know how to use the 3D printer? He was like, ``here is the project, let's do it.'' And anything that I was unsure of, I would just be like, how do I do this, and he would just accept it and teach me like the 3D printer. I did not know how to use a 3D printer. Now I do... I learned about using 3D printers, which is a lot of fun... He does not question, what you can do, but instead leaves it open-ended for you to come in when you need help.'' This emphasizes how mentors can build a solid and trusting relationship with their students. In the extensive study \cite{zohrabi_alaee_impact_2022}, we argued how this communicative relation between mentor and mentee could impact mentees' self-efficacy, sense of belonging, and physics identity, which lead to a positive view on their future career. 
In short, our data shows that many participants described this experience as a ``real'' research practice, where they achieved their goals (research goals and personal goals for participating in this program). The remote format has not opened a gap between what mentees wanted from the REU program and what they got at the end of the program. Interestingly, our findings show that the remote research experience was closely equivalent to traditional in-person REU experience in gaining new knowledge, applying to graduate school, and gaining different skills. It is important to remember that most of our mentees had a previous in-person REU experience to compare their remote experience.

\section{Conclusion and implications}
As Engestr{\"o}m argues, identifying different components of any activity system can help us to understand challenges and barriers within that system, which lead to refining a better system \cite{engestrom_learning_1987, Engestrom2001, engestrom_learning_2014}. %Beyond the world facing a health crisis like COVID-19 in the summer of 2020, learning about what happened in a complex REU activity could help us understand how research can be done remotely and the possible benefits and outcomes of doing research in the remote environment.

Our first research question asked: \textit{What are some tools, communities, norms, and division of labors that make up the research experience?} Our data shows a holistic perspective of the remote REU program and how mentees used tools and communities to engage in the remote learning space in order to achieve their goals. The second research question was: \textit{What are some of the challenges of the remote REU program?} Based on our semi-structured interviews, mentees shared with us different types of challenges they faced during the remote REU program. %However, the most repeated challenge was working at home and technical difficulties, leading to frustrating situations. 
As we outline in Section \ref{Sec: Challenges rq2}, COVID-19 and the remote format impacted students' learning. For instance, essential components of effective learning, such as easy availability of helpful feedback and having a good mentor-mentee relationship, were harder in a remote format. All students experienced a new life in quarantine. Many of them expressed that they had to set up a workspace in different rooms of the house or a basement to be relatively free from distractions while working. Thus, these findings are in tandem with the results from previous studies around learning challenges in a remote format \cite{2007AGUFMED33B1226H, gonzalez2021emergency, nambiar2020impact, zhang2020suspending}. In addition, the limited exposure of REU students to the REU community was more challenging. Although each REU program had different social events such as meetings with previous REUs, check-in meetings, group presentations, movie nights, seminars, and elective courses to build community among their REU students, it was still not enough. Most mentees enjoyed the social events, but they would have liked more whole-group interactions. 

%Besides, some mentees talked about small aspects of the remote program. For instance, Emma stated she wanted REU T-shirts or commemorative mug from this experience and his mentor gave her couple T-shirts from some of the art to science activities he did in the past. Andrew said he ``got nervous'' to ask for help remotely at the beginning of the program, but ``I normally come around and do it anyways but like just getting in the habit of being able to get on Slack and ask someone a really simple question and being confident in my ability to do so, I guess.'' 
%Some of them expressed after attending the remote research experience, they gained insight into graduate school and were able to narrow down their career options.

Our last research question was: \textit{What are some of the outcomes of the remote REU program?} Our data show that the outcomes of the remote REU program is not too far from the generally accepted benefits for any undergraduate research experience. Despite the challenges of the remote REU experience, most mentees ($N$=9) gained clearer ideas about their future career goals. %Furthermore, converting experimental experiences to remote research experiences was beneficial for some mentors to try their previous or future hands-on research more computational. %Besides, other mentors, like Emma and Bruce's mentors, figured out ways for mentees to develop hands-on activities at home in addition to experience computational work.
%This study explores some of the students' outcomes throughout the remote research program. 

Although we attempted to lighten many of the aspects associated with remote research experience in our study, there is a need for some future work to collect same data and compare students who have access to the remote lab and other students who involve in the in-person lab without discrimination. Second, to what extent do remote research programs allow students to engage in research? Future studies need to seek more information on the underlying mechanisms, including how and why these outcomes happened, which mediators (e.g., norms, community, tools) support particular outcomes, and how these short-term outcomes increase long-term retention in STEM majors. Third, future research is needed to examine how remote research experience impacts different populations, such as participation from marginalized groups. Lastly, in examining the benefit of the remote research programs more research is needed to uncover more specific practices and support structures that help students develop the skills necessary to succeed in their future career. 

Although COVID-19 has created some challenges for REU participants and their mentors, various opportunities are also evolved. For example, research mentors and REU coordinators may understand how research can be done remotely now and the possible benefits and outcomes of doing research in a remote environment. They could design strategies and methods based on participants' needs in a way that has a more long-term impact on the student's retention in STEM fields.

%Beyond the world facing a health crisis like COVID-19 in the summer of 2020, learning about what happened in a complex REU activity could help us understand how research can be done remotely and the possible benefits and outcomes of doing research in the remote environment. For instance, 
The findings of this research provide evidence that the remote REU experience was beneficial to students and should be considered a ``real'' research experience. Hence, there is an opportunity to develop and expand remote research programs in order to make scientific research and its benefits more accessible. Remote research experiences could support inclusion in STEM by benefiting students who have travel restrictions due to health conditions, family obligations, or financial challenges. However, though the resource demands on students (dedicated physical space, a computer, and high speed internet) would need to be carefully addressed. Furthermore, remote research may have unique benefits, such as increasing students' sense of confidence when they overcome challenges on their own. While COVID-19 was an unwelcome crisis, it has caused us to rethink our standard practices and given insight for future improvements in undergraduate research opportunities.

\section{ACKNOWLEDGEMENT}
The authors would like to thank all individuals who shared their REU experiences. Without their participation this work would not have been possible. This work was supported by the National Science Foundation under Grant No. 1846321.
\bibliographystyle{apsrev4-1}
\bibliography{RQ1.bib}

%merlin.mbs apsrev4-1.bst 2010-07-25 4.21a (PWD, AO, DPC) hacked
%Control: key (0)
%Control: author (72) initials jnrlst
%Control: editor formatted (1) identically to author
%Control: production of article title (-1) disabled
%Control: page (0) single
%Control: year (1) truncated
%Control: production of eprint (0) enabled
\begin{thebibliography}{53}%
\makeatletter
\providecommand \@ifxundefined [1]{%
 \@ifx{#1\undefined}
}%
\providecommand \@ifnum [1]{%
 \ifnum #1\expandafter \@firstoftwo
 \else \expandafter \@secondoftwo
 \fi
}%
\providecommand \@ifx [1]{%
 \ifx #1\expandafter \@firstoftwo
 \else \expandafter \@secondoftwo
 \fi
}%
\providecommand \natexlab [1]{#1}%
\providecommand \enquote  [1]{``#1''}%
\providecommand \bibnamefont  [1]{#1}%
\providecommand \bibfnamefont [1]{#1}%
\providecommand \citenamefont [1]{#1}%
\providecommand \href@noop [0]{\@secondoftwo}%
\providecommand \href [0]{\begingroup \@sanitize@url \@href}%
\providecommand \@href[1]{\@@startlink{#1}\@@href}%
\providecommand \@@href[1]{\endgroup#1\@@endlink}%
\providecommand \@sanitize@url [0]{\catcode `\\12\catcode `\$12\catcode
  `\&12\catcode `\#12\catcode `\^12\catcode `\_12\catcode `\%12\relax}%
\providecommand \@@startlink[1]{}%
\providecommand \@@endlink[0]{}%
\providecommand \url  [0]{\begingroup\@sanitize@url \@url }%
\providecommand \@url [1]{\endgroup\@href {#1}{\urlprefix }}%
\providecommand \urlprefix  [0]{URL }%
\providecommand \Eprint [0]{\href }%
\providecommand \doibase [0]{http://dx.doi.org/}%
\providecommand \selectlanguage [0]{\@gobble}%
\providecommand \bibinfo  [0]{\@secondoftwo}%
\providecommand \bibfield  [0]{\@secondoftwo}%
\providecommand \translation [1]{[#1]}%
\providecommand \BibitemOpen [0]{}%
\providecommand \bibitemStop [0]{}%
\providecommand \bibitemNoStop [0]{.\EOS\space}%
\providecommand \EOS [0]{\spacefactor3000\relax}%
\providecommand \BibitemShut  [1]{\csname bibitem#1\endcsname}%
\let\auto@bib@innerbib\@empty
%</preamble>
\bibitem [{\citenamefont {Council}(1999)}]{council_transforming_1999}%
  \BibitemOpen
  \bibfield  {author} {\bibinfo {author} {\bibfnamefont {N.~R.}\ \bibnamefont
  {Council}},\ }\href@noop {} {\emph {\bibinfo {title} {Transforming
  Undergraduate Education in Science, Mathematics, Engineering, and
  Technology}}}\ (\bibinfo  {publisher} {The National Academies Press},\
  \bibinfo {address} {Washington, DC},\ \bibinfo {year} {1999})\BibitemShut
  {NoStop}%
\bibitem [{\citenamefont {Council}(2002)}]{council_improving_2002}%
  \BibitemOpen
  \bibfield  {author} {\bibinfo {author} {\bibfnamefont {N.~R.}\ \bibnamefont
  {Council}},\ }\href {\doibase https://doi.org/10.17226/10711} {\emph
  {\bibinfo {title} {Improving undergraduate instruction in Science,
  Technology, Engineering, and Mathematics: report of a workshop}}}\ (\bibinfo
  {publisher} {The National Academies Press},\ \bibinfo {address} {Washington,
  DC},\ \bibinfo {year} {2002})\BibitemShut {NoStop}%
\bibitem [{\citenamefont {Wenzel}\ and\ \citenamefont
  {Karukstis}(2004)}]{wenzel_enhancing_2004}%
  \BibitemOpen
  \bibfield  {author} {\bibinfo {author} {\bibfnamefont {T.~J.}\ \bibnamefont
  {Wenzel}}\ and\ \bibinfo {author} {\bibfnamefont {K.~K.}\ \bibnamefont
  {Karukstis}},\ }\href {\doibase https://doi.org/10.1021/ed081p468} {\bibfield
   {journal} {\bibinfo  {journal} {J. Chem. Educ.}\ }\textbf {\bibinfo {volume}
  {81}},\ \bibinfo {pages} {468} (\bibinfo {year} {2004})},\ \bibinfo {note}
  {publisher: American Chemical Society}\BibitemShut {NoStop}%
\bibitem [{\citenamefont {Kenny}\ \emph {et~al.}(2001)\citenamefont {Kenny},
  \citenamefont {Thomas}, \citenamefont {Katkin}, \citenamefont {Lemming},
  \citenamefont {Smith}, \citenamefont {Glaser},\ and\ \citenamefont
  {Gross}}]{Kenny_2001}%
  \BibitemOpen
  \bibfield  {author} {\bibinfo {author} {\bibfnamefont {S.~S.}\ \bibnamefont
  {Kenny}}, \bibinfo {author} {\bibfnamefont {E.}~\bibnamefont {Thomas}},
  \bibinfo {author} {\bibfnamefont {W.}~\bibnamefont {Katkin}}, \bibinfo
  {author} {\bibfnamefont {M.}~\bibnamefont {Lemming}}, \bibinfo {author}
  {\bibfnamefont {P.}~\bibnamefont {Smith}}, \bibinfo {author} {\bibfnamefont
  {M.}~\bibnamefont {Glaser}}, \ and\ \bibinfo {author} {\bibfnamefont
  {W.}~\bibnamefont {Gross}},\ }\href@noop {} {\emph {\bibinfo {title}
  {Reinventing undergraduate education: three years after the Boyer report}}}\
  (\bibinfo  {publisher} {Stony Brook University: Office of the President},\
  \bibinfo {address} {NY},\ \bibinfo {year} {2001})\BibitemShut {NoStop}%
\bibitem [{\citenamefont {Kuh}(2008)}]{kuh_high-impact_2008}%
  \BibitemOpen
  \bibfield  {author} {\bibinfo {author} {\bibfnamefont {G.~D.}\ \bibnamefont
  {Kuh}},\ }\href@noop {} {\emph {\bibinfo {title} {High-impact educational
  practices: what they are, who has access to them, and why they matter}}}\
  (\bibinfo  {publisher} {Association of American Colleges and Universities},\
  \bibinfo {address} {Washington, D.C.},\ \bibinfo {year} {2008})\BibitemShut
  {NoStop}%
\bibitem [{\citenamefont {Kardash}(2000)}]{kardash_evaluation_2000}%
  \BibitemOpen
  \bibfield  {author} {\bibinfo {author} {\bibfnamefont {C.~M.}\ \bibnamefont
  {Kardash}},\ }\href {\doibase https://doi.org/10.1037/0022-0663.92.1.191}
  {\bibfield  {journal} {\bibinfo  {journal} {J. Educ. Psychol.}\ }\textbf
  {\bibinfo {volume} {92}},\ \bibinfo {pages} {191} (\bibinfo {year}
  {2000})}\BibitemShut {NoStop}%
\bibitem [{\citenamefont {Lopatto}(2007)}]{lopatto_undergraduate_2007}%
  \BibitemOpen
  \bibfield  {author} {\bibinfo {author} {\bibfnamefont {D.}~\bibnamefont
  {Lopatto}},\ }\href {\doibase https://doi.org/10.1187/cbe.07-06-0039}
  {\bibfield  {journal} {\bibinfo  {journal} {CBE Life Sci. Educ.}\ }\textbf
  {\bibinfo {volume} {6}},\ \bibinfo {pages} {297} (\bibinfo {year}
  {2007})}\BibitemShut {NoStop}%
\bibitem [{\citenamefont {Hathaway}\ \emph {et~al.}(2002)\citenamefont
  {Hathaway}, \citenamefont {Nagda},\ and\ \citenamefont
  {Gregerman}}]{hathaway_relationship_2002}%
  \BibitemOpen
  \bibfield  {author} {\bibinfo {author} {\bibfnamefont {R.~S.}\ \bibnamefont
  {Hathaway}}, \bibinfo {author} {\bibfnamefont {B.~R.~A.}\ \bibnamefont
  {Nagda}}, \ and\ \bibinfo {author} {\bibfnamefont {S.~R.}\ \bibnamefont
  {Gregerman}},\ }\href@noop {} {\bibfield  {journal} {\bibinfo  {journal} {J.
  Coll. Stud. Dev.}\ }\textbf {\bibinfo {volume} {43}},\ \bibinfo {pages} {614}
  (\bibinfo {year} {2002})}\BibitemShut {NoStop}%
\bibitem [{\citenamefont {Seymour}\ \emph {et~al.}(2004)\citenamefont
  {Seymour}, \citenamefont {Hunter}, \citenamefont {Laursen},\ and\
  \citenamefont {DeAntoni}}]{seymour_establishing_2004}%
  \BibitemOpen
  \bibfield  {author} {\bibinfo {author} {\bibfnamefont {E.}~\bibnamefont
  {Seymour}}, \bibinfo {author} {\bibfnamefont {A.-B.}\ \bibnamefont {Hunter}},
  \bibinfo {author} {\bibfnamefont {S.~L.}\ \bibnamefont {Laursen}}, \ and\
  \bibinfo {author} {\bibfnamefont {T.}~\bibnamefont {DeAntoni}},\ }\href@noop
  {} {\bibfield  {journal} {\bibinfo  {journal} {Sci. Educ.}\ }\textbf
  {\bibinfo {volume} {88}} (\bibinfo {year} {2004})}\BibitemShut {NoStop}%
\bibitem [{\citenamefont {Lopatto}(2004)}]{lopatto_survey_2004}%
  \BibitemOpen
  \bibfield  {author} {\bibinfo {author} {\bibfnamefont {D.}~\bibnamefont
  {Lopatto}},\ }\href {\doibase https://doi.org/10.1187/cbe.04-07-0045}
  {\bibfield  {journal} {\bibinfo  {journal} {CBE Life Sci. Educ.}\ }\textbf
  {\bibinfo {volume} {3}},\ \bibinfo {pages} {270} (\bibinfo {year}
  {2004})}\BibitemShut {NoStop}%
\bibitem [{\citenamefont {Hunter}\ \emph {et~al.}(2007)\citenamefont {Hunter},
  \citenamefont {Laursen},\ and\ \citenamefont
  {Seymour}}]{hunter_becoming_2007}%
  \BibitemOpen
  \bibfield  {author} {\bibinfo {author} {\bibfnamefont {A.-B.}\ \bibnamefont
  {Hunter}}, \bibinfo {author} {\bibfnamefont {S.~L.}\ \bibnamefont {Laursen}},
  \ and\ \bibinfo {author} {\bibfnamefont {E.}~\bibnamefont {Seymour}},\
  }\href@noop {} {\bibfield  {journal} {\bibinfo  {journal} {Sci. Educ.}\
  }\textbf {\bibinfo {volume} {91}},\ \bibinfo {pages} {36} (\bibinfo {year}
  {2007})}\BibitemShut {NoStop}%
\bibitem [{\citenamefont {Laursen}\ \emph {et~al.}(2010)\citenamefont
  {Laursen}, \citenamefont {Hunter}, \citenamefont {Seymour}, \citenamefont
  {Thiry},\ and\ \citenamefont {Melton}}]{laursen_undergraduate_2010}%
  \BibitemOpen
  \bibfield  {author} {\bibinfo {author} {\bibfnamefont {S.}~\bibnamefont
  {Laursen}}, \bibinfo {author} {\bibfnamefont {A.-B.}\ \bibnamefont {Hunter}},
  \bibinfo {author} {\bibfnamefont {E.}~\bibnamefont {Seymour}}, \bibinfo
  {author} {\bibfnamefont {H.}~\bibnamefont {Thiry}}, \ and\ \bibinfo {author}
  {\bibfnamefont {G.}~\bibnamefont {Melton}},\ }\href@noop {} {\emph {\bibinfo
  {title} {Undergraduate research in the sciences: engaging students in real
  science}}}\ (\bibinfo  {publisher} {Jossey-Bass},\ \bibinfo {year}
  {2010})\BibitemShut {NoStop}%
\bibitem [{\citenamefont {Estrada}\ \emph {et~al.}(2011)\citenamefont
  {Estrada}, \citenamefont {Woodcock}, \citenamefont {Hernandez},\ and\
  \citenamefont {Schultz}}]{estrada_toward_2011}%
  \BibitemOpen
  \bibfield  {author} {\bibinfo {author} {\bibfnamefont {M.}~\bibnamefont
  {Estrada}}, \bibinfo {author} {\bibfnamefont {A.}~\bibnamefont {Woodcock}},
  \bibinfo {author} {\bibfnamefont {P.~R.}\ \bibnamefont {Hernandez}}, \ and\
  \bibinfo {author} {\bibfnamefont {P.~W.}\ \bibnamefont {Schultz}},\ }\href
  {\doibase https://doi.org/10.1037/a0020743} {\bibfield  {journal} {\bibinfo
  {journal} {J. Educ. Psychol.}\ }\textbf {\bibinfo {volume} {103}},\ \bibinfo
  {pages} {206} (\bibinfo {year} {2011})}\BibitemShut {NoStop}%
\bibitem [{\citenamefont {Graham}\ \emph {et~al.}(2013)\citenamefont {Graham},
  \citenamefont {Frederick}, \citenamefont {Byars-Winston}, \citenamefont
  {Hunter},\ and\ \citenamefont {Handelsman}}]{graham_increasing_2013}%
  \BibitemOpen
  \bibfield  {author} {\bibinfo {author} {\bibfnamefont {M.~J.}\ \bibnamefont
  {Graham}}, \bibinfo {author} {\bibfnamefont {J.}~\bibnamefont {Frederick}},
  \bibinfo {author} {\bibfnamefont {A.}~\bibnamefont {Byars-Winston}}, \bibinfo
  {author} {\bibfnamefont {A.-B.}\ \bibnamefont {Hunter}}, \ and\ \bibinfo
  {author} {\bibfnamefont {J.}~\bibnamefont {Handelsman}},\ }\href {\doibase
  https://doi.org/10.1126/science.1240487} {\bibfield  {journal} {\bibinfo
  {journal} {Science}\ }\textbf {\bibinfo {volume} {341}},\ \bibinfo {pages}
  {1455} (\bibinfo {year} {2013})}\BibitemShut {NoStop}%
\bibitem [{\citenamefont {Hausmann}\ \emph {et~al.}(2007)\citenamefont
  {Hausmann}, \citenamefont {Schofield},\ and\ \citenamefont
  {Woods}}]{hausmann_sense_2007}%
  \BibitemOpen
  \bibfield  {author} {\bibinfo {author} {\bibfnamefont {L.~R.~M.}\
  \bibnamefont {Hausmann}}, \bibinfo {author} {\bibfnamefont {J.~W.}\
  \bibnamefont {Schofield}}, \ and\ \bibinfo {author} {\bibfnamefont {R.~L.}\
  \bibnamefont {Woods}},\ }\href@noop {} {\bibfield  {journal} {\bibinfo
  {journal} {Res. High. Educ.}\ }\textbf {\bibinfo {volume} {48}},\ \bibinfo
  {pages} {803} (\bibinfo {year} {2007})}\BibitemShut {NoStop}%
\bibitem [{\citenamefont {Dolan}\ and\ \citenamefont
  {Johnson}(2009)}]{dolan_toward_2009}%
  \BibitemOpen
  \bibfield  {author} {\bibinfo {author} {\bibfnamefont {E.}~\bibnamefont
  {Dolan}}\ and\ \bibinfo {author} {\bibfnamefont {D.}~\bibnamefont
  {Johnson}},\ }\href {\doibase https://doi.org/10.1007/s10956-009-9165-3}
  {\bibfield  {journal} {\bibinfo  {journal} {J. Sci. Educ. Technol.}\ }\textbf
  {\bibinfo {volume} {18}},\ \bibinfo {pages} {487} (\bibinfo {year}
  {2009})}\BibitemShut {NoStop}%
\bibitem [{\citenamefont {Eagan}\ \emph {et~al.}(2013)\citenamefont {Eagan},
  \citenamefont {Hurtado}, \citenamefont {Chang}, \citenamefont {Garcia},
  \citenamefont {Herrera},\ and\ \citenamefont {Garibay}}]{eagan_making_2013}%
  \BibitemOpen
  \bibfield  {author} {\bibinfo {author} {\bibfnamefont {M.~K.}\ \bibnamefont
  {Eagan}}, \bibinfo {author} {\bibfnamefont {S.}~\bibnamefont {Hurtado}},
  \bibinfo {author} {\bibfnamefont {M.~J.}\ \bibnamefont {Chang}}, \bibinfo
  {author} {\bibfnamefont {G.~A.}\ \bibnamefont {Garcia}}, \bibinfo {author}
  {\bibfnamefont {F.~A.}\ \bibnamefont {Herrera}}, \ and\ \bibinfo {author}
  {\bibfnamefont {J.~C.}\ \bibnamefont {Garibay}},\ }\href {\doibase
  https://doi.org/10.3102/0002831213482038} {\bibfield  {journal} {\bibinfo
  {journal} {Am. Educ. Res. J.}\ }\textbf {\bibinfo {volume} {50}},\ \bibinfo
  {pages} {683} (\bibinfo {year} {2013})}\BibitemShut {NoStop}%
\bibitem [{\citenamefont {Zohrabi-Alaee}\ \emph {et~al.}(2022)\citenamefont
  {Zohrabi-Alaee}, \citenamefont {Campbell},\ and\ \citenamefont
  {Zwickl}}]{zohrabi_alaee_impact_2022}%
  \BibitemOpen
  \bibfield  {author} {\bibinfo {author} {\bibfnamefont {D.}~\bibnamefont
  {Zohrabi-Alaee}}, \bibinfo {author} {\bibfnamefont {M.~K.}\ \bibnamefont
  {Campbell}}, \ and\ \bibinfo {author} {\bibfnamefont {B.~M.}\ \bibnamefont
  {Zwickl}},\ }\href@noop {} {\bibfield  {journal} {\bibinfo  {journal} {Phys.
  Rev. Phys. Educ. Res.}\ }\textbf {\bibinfo {volume} {18}},\ \bibinfo {pages}
  {010101} (\bibinfo {year} {2022})}\BibitemShut {NoStop}%
\bibitem [{\citenamefont {of~Sciences}\ \emph {et~al.}(2017)\citenamefont
  {of~Sciences}, \citenamefont {Medicine},\ and\ \citenamefont
  {Engineering}}]{national_academies_of_sciences_undergraduate_2017}%
  \BibitemOpen
  \bibfield  {author} {\bibinfo {author} {\bibfnamefont {N.~A.}\ \bibnamefont
  {of~Sciences}}, \bibinfo {author} {\bibnamefont {Medicine}}, \ and\ \bibinfo
  {author} {\bibnamefont {Engineering}},\ }\href@noop {} {\emph {\bibinfo
  {title} {Undergraduate Research Experiences for STEM students: successes,
  challenges, and opportunities}}},\ edited by\ \bibinfo {editor}
  {\bibfnamefont {J.}~\bibnamefont {Gentile}}, \bibinfo {editor} {\bibfnamefont
  {K.}~\bibnamefont {Brenner}}, \ and\ \bibinfo {editor} {\bibfnamefont
  {A.}~\bibnamefont {Stephens}}\ (\bibinfo  {publisher} {The National Academies
  Press},\ \bibinfo {address} {Washington, DC},\ \bibinfo {year}
  {2017})\BibitemShut {NoStop}%
\bibitem [{\citenamefont {Blockus}(2016)}]{blockus_strengthening_2016}%
  \BibitemOpen
  \bibfield  {author} {\bibinfo {author} {\bibfnamefont {L.}~\bibnamefont
  {Blockus}},\ }\href@noop {} {\bibfield  {journal} {\bibinfo  {journal}
  {National Academies of Sciences, Engineering, and Medicine (NASEM)}\ ,\
  \bibinfo {pages} {31}} (\bibinfo {year} {2016})}\BibitemShut {NoStop}%
\bibitem [{\citenamefont {Russell}\ \emph {et~al.}(2007)\citenamefont
  {Russell}, \citenamefont {Hancock},\ and\ \citenamefont
  {McCullough}}]{russell_benefits_2007}%
  \BibitemOpen
  \bibfield  {author} {\bibinfo {author} {\bibfnamefont {S.~H.}\ \bibnamefont
  {Russell}}, \bibinfo {author} {\bibfnamefont {M.~P.}\ \bibnamefont
  {Hancock}}, \ and\ \bibinfo {author} {\bibfnamefont {J.}~\bibnamefont
  {McCullough}},\ }\href@noop {} {\bibfield  {journal} {\bibinfo  {journal}
  {Science}\ }\textbf {\bibinfo {volume} {316}},\ \bibinfo {pages} {548}
  (\bibinfo {year} {2007})}\BibitemShut {NoStop}%
\bibitem [{\citenamefont {Forrester}(2021)}]{forrester_how_2021}%
  \BibitemOpen
  \bibfield  {author} {\bibinfo {author} {\bibfnamefont {N.}~\bibnamefont
  {Forrester}},\ }\href {\doibase DOI: 10.1038/d41586-021-01209-2} {\bibfield
  {journal} {\bibinfo  {journal} {Nature}\ } (\bibinfo {year} {2021}),\ DOI:
  10.1038/d41586-021-01209-2}\BibitemShut {NoStop}%
\bibitem [{\citenamefont {Leont'ev}(1974)}]{leontev_problem_1974}%
  \BibitemOpen
  \bibfield  {author} {\bibinfo {author} {\bibfnamefont {A.~N.}\ \bibnamefont
  {Leont'ev}},\ }\href@noop {} {\bibfield  {journal} {\bibinfo  {journal}
  {Soviet Psychology}\ }\textbf {\bibinfo {volume} {13}},\ \bibinfo {pages} {4}
  (\bibinfo {year} {1974})},\ \bibinfo {note} {publisher:
  Routledge}\BibitemShut {NoStop}%
\bibitem [{\citenamefont {Roth}\ and\ \citenamefont
  {Lee}(2007)}]{roth_vygotskys_2007}%
  \BibitemOpen
  \bibfield  {author} {\bibinfo {author} {\bibfnamefont {W.-M.}\ \bibnamefont
  {Roth}}\ and\ \bibinfo {author} {\bibfnamefont {Y.-J.}\ \bibnamefont {Lee}},\
  }\href@noop {} {\bibfield  {journal} {\bibinfo  {journal} {Rev. Educ. Res.}\
  }\textbf {\bibinfo {volume} {77}},\ \bibinfo {pages} {186} (\bibinfo {year}
  {2007})},\ \bibinfo {note} {publisher: American Educational Research
  Association}\BibitemShut {NoStop}%
\bibitem [{\citenamefont {Engestr{\"o}m}(1987)}]{engestrom_learning_1987}%
  \BibitemOpen
  \bibfield  {author} {\bibinfo {author} {\bibfnamefont {Y.}~\bibnamefont
  {Engestr{\"o}m}},\ }\href@noop {} {\emph {\bibinfo {title} {Learning by
  expanding: A activity-theoretical approach to developmental research}}}\
  (\bibinfo  {publisher} {Cambridge University Press},\ \bibinfo {address} {New
  York, NY},\ \bibinfo {year} {1987})\BibitemShut {NoStop}%
\bibitem [{\citenamefont {Vygotsky}\ \emph {et~al.}(1978)\citenamefont
  {Vygotsky}, \citenamefont {Cole}, \citenamefont {John-Steiner}, \citenamefont
  {Scribner},\ and\ \citenamefont {Souberman}}]{vygotsky_mind_1978}%
  \BibitemOpen
  \bibfield  {author} {\bibinfo {author} {\bibfnamefont {L.~S.}\ \bibnamefont
  {Vygotsky}}, \bibinfo {author} {\bibfnamefont {M.}~\bibnamefont {Cole}},
  \bibinfo {author} {\bibfnamefont {V.}~\bibnamefont {John-Steiner}}, \bibinfo
  {author} {\bibfnamefont {S.}~\bibnamefont {Scribner}}, \ and\ \bibinfo
  {author} {\bibfnamefont {E.}~\bibnamefont {Souberman}},\ }\href@noop {}
  {\emph {\bibinfo {title} {Mind in society: development of higher
  psychological processes}}}\ (\bibinfo  {publisher} {Harvard University
  Press},\ \bibinfo {year} {1978})\BibitemShut {NoStop}%
\bibitem [{\citenamefont {Engestr{\"o}m}(2014)}]{engestrom_learning_2014}%
  \BibitemOpen
  \bibfield  {author} {\bibinfo {author} {\bibfnamefont {Y.}~\bibnamefont
  {Engestr{\"o}m}},\ }\href {\doibase 10.1017/CBO9781139814744} {\emph
  {\bibinfo {title} {Learning by Expanding: An Activity-Theoretical Approach to
  Developmental Research}}}\ (\bibinfo  {publisher} {Cambridge University
  Press},\ \bibinfo {address} {Cambridge},\ \bibinfo {year} {2014})\BibitemShut
  {NoStop}%
\bibitem [{\citenamefont {Engestr{\"o}m}(1999)}]{engestrom_expansive_1999}%
  \BibitemOpen
  \bibfield  {author} {\bibinfo {author} {\bibfnamefont {Y.}~\bibnamefont
  {Engestr{\"o}m}},\ }\href@noop {} {\bibfield  {journal} {\bibinfo  {journal}
  {Comput. Support. Coop. Work.}\ }\textbf {\bibinfo {volume} {8}},\ \bibinfo
  {pages} {63} (\bibinfo {year} {1999})}\BibitemShut {NoStop}%
\bibitem [{\citenamefont {Yamagata-Lynch}(2010)}]{yamagata2010activity}%
  \BibitemOpen
  \bibfield  {author} {\bibinfo {author} {\bibfnamefont {L.~C.}\ \bibnamefont
  {Yamagata-Lynch}},\ }\href@noop {} {\emph {\bibinfo {title} {Activity systems
  analysis methods: Understanding complex learning environments}}}\ (\bibinfo
  {publisher} {Springer Science \& Business Media},\ \bibinfo {year}
  {2010})\BibitemShut {NoStop}%
\bibitem [{\citenamefont {Helba}\ \emph {et~al.}(2019)\citenamefont {Helba},
  \citenamefont {Porter}, \citenamefont {Nicholson},\ and\ \citenamefont
  {Ivie}}]{AIP2019}%
  \BibitemOpen
  \bibfield  {author} {\bibinfo {author} {\bibfnamefont {C.}~\bibnamefont
  {Helba}}, \bibinfo {author} {\bibfnamefont {A.~M.}\ \bibnamefont {Porter}},
  \bibinfo {author} {\bibfnamefont {S.}~\bibnamefont {Nicholson}}, \ and\
  \bibinfo {author} {\bibfnamefont {R.}~\bibnamefont {Ivie}},\ }\href
  {https://www.aip.org/statistics/reports/women-among-physics-and-astronomy-faculty}
  {\enquote {\bibinfo {title} {Women among physics and astronomy faculty
  results from the 2018 academic workforce survey},}\ } (\bibinfo {year}
  {2019})\BibitemShut {NoStop}%
\bibitem [{\citenamefont {Walsh}\ and\ \citenamefont {Tyler}(2022)}]{AIP2022}%
  \BibitemOpen
  \bibfield  {author} {\bibinfo {author} {\bibfnamefont {C.}~\bibnamefont
  {Walsh}}\ and\ \bibinfo {author} {\bibfnamefont {J.}~\bibnamefont {Tyler}},\
  }\href {https://www.aip.org/statistics/reports/covid-faculty-qualityofwork}
  {\enquote {\bibinfo {title} {Self-reported changes in quality of work as a
  result of the covid-19 pandemic for faculty members in physics and
  astronomy},}\ } (\bibinfo {year} {2022})\BibitemShut {NoStop}%
\bibitem [{\citenamefont {Myers}\ \emph {et~al.}(2020)\citenamefont {Myers},
  \citenamefont {Tham}, \citenamefont {Yin}, \citenamefont {Cohodes},
  \citenamefont {Thursby}, \citenamefont {Thursby}, \citenamefont {Schiffer},
  \citenamefont {Walsh}, \citenamefont {Lakhani},\ and\ \citenamefont
  {Wang}}]{myers2020unequal}%
  \BibitemOpen
  \bibfield  {author} {\bibinfo {author} {\bibfnamefont {K.~R.}\ \bibnamefont
  {Myers}}, \bibinfo {author} {\bibfnamefont {W.~Y.}\ \bibnamefont {Tham}},
  \bibinfo {author} {\bibfnamefont {Y.}~\bibnamefont {Yin}}, \bibinfo {author}
  {\bibfnamefont {N.}~\bibnamefont {Cohodes}}, \bibinfo {author} {\bibfnamefont
  {J.~G.}\ \bibnamefont {Thursby}}, \bibinfo {author} {\bibfnamefont {M.~C.}\
  \bibnamefont {Thursby}}, \bibinfo {author} {\bibfnamefont {P.}~\bibnamefont
  {Schiffer}}, \bibinfo {author} {\bibfnamefont {J.~T.}\ \bibnamefont {Walsh}},
  \bibinfo {author} {\bibfnamefont {K.~R.}\ \bibnamefont {Lakhani}}, \ and\
  \bibinfo {author} {\bibfnamefont {D.}~\bibnamefont {Wang}},\ }\href@noop {}
  {\bibfield  {journal} {\bibinfo  {journal} {Nat. Hum. Behav.}\ }\textbf
  {\bibinfo {volume} {4}},\ \bibinfo {pages} {880} (\bibinfo {year}
  {2020})}\BibitemShut {NoStop}%
\bibitem [{\citenamefont {Krukowski}\ \emph {et~al.}(2021)\citenamefont
  {Krukowski}, \citenamefont {Jagsi},\ and\ \citenamefont
  {Cardel}}]{katz2021re}%
  \BibitemOpen
  \bibfield  {author} {\bibinfo {author} {\bibfnamefont {R.}~\bibnamefont
  {Krukowski}}, \bibinfo {author} {\bibfnamefont {R.}~\bibnamefont {Jagsi}}, \
  and\ \bibinfo {author} {\bibfnamefont {M.}~\bibnamefont {Cardel}},\
  }\href@noop {} {\bibfield  {journal} {\bibinfo  {journal} {J. Womens Health}\
  }\textbf {\bibinfo {volume} {30}},\ \bibinfo {pages} {341} (\bibinfo {year}
  {2021})}\BibitemShut {NoStop}%
\bibitem [{\citenamefont {Marton}\ and\ \citenamefont
  {Booth}(1997)}]{Marton1997}%
  \BibitemOpen
  \bibfield  {author} {\bibinfo {author} {\bibfnamefont {F.}~\bibnamefont
  {Marton}}\ and\ \bibinfo {author} {\bibfnamefont {S.~A.}\ \bibnamefont
  {Booth}},\ }\href
  {http://gateway.library.qut.edu.au/login?url=https://www.taylorfrancis.com/books/9781136495762}
  {\emph {\bibinfo {title} {Learning and awareness}}}\ (\bibinfo  {publisher}
  {Mahwah, N.J. : L. Erlbaum Associates},\ \bibinfo {year} {1997})\BibitemShut
  {NoStop}%
\bibitem [{\citenamefont {Marton}(2000)}]{marton2000structure}%
  \BibitemOpen
  \bibfield  {author} {\bibinfo {author} {\bibfnamefont {F.}~\bibnamefont
  {Marton}},\ }\href@noop {} {\bibfield  {journal} {\bibinfo  {journal}
  {Phenomenography}\ ,\ \bibinfo {pages} {102}} (\bibinfo {year}
  {2000})}\BibitemShut {NoStop}%
\bibitem [{\citenamefont {Deddose}(2018)}]{noauthor_dedoose_2018}%
  \BibitemOpen
  \bibfield  {author} {\bibinfo {author} {\bibnamefont {Deddose}},\ }\href
  {https://app.dedoose.com/App/?Version=8.0.35} {\enquote {\bibinfo {title}
  {Dedoose version 8.0.35, web application for managing, analyzing, and
  presenting qualitative and mixed method research data},}\ } (\bibinfo {year}
  {2018})\BibitemShut {NoStop}%
\bibitem [{\citenamefont {Zhao}(2007)}]{zhao2007cultural}%
  \BibitemOpen
  \bibfield  {author} {\bibinfo {author} {\bibfnamefont {Y.}~\bibnamefont
  {Zhao}},\ }\href@noop {} {\bibfield  {journal} {\bibinfo  {journal}
  {Intercultural Communication Studies}\ }\textbf {\bibinfo {volume} {16}},\
  \bibinfo {pages} {129} (\bibinfo {year} {2007})}\BibitemShut {NoStop}%
\bibitem [{\citenamefont {Rachel~Zhou}\ \emph {et~al.}(2005)\citenamefont
  {Rachel~Zhou}, \citenamefont {Knoke},\ and\ \citenamefont
  {Sakamoto}}]{rachel2005rethinking}%
  \BibitemOpen
  \bibfield  {author} {\bibinfo {author} {\bibfnamefont {Y.}~\bibnamefont
  {Rachel~Zhou}}, \bibinfo {author} {\bibfnamefont {D.}~\bibnamefont {Knoke}},
  \ and\ \bibinfo {author} {\bibfnamefont {I.}~\bibnamefont {Sakamoto}},\
  }\href@noop {} {\bibfield  {journal} {\bibinfo  {journal} {International
  Journal of Inclusive Education}\ }\textbf {\bibinfo {volume} {9}},\ \bibinfo
  {pages} {287} (\bibinfo {year} {2005})}\BibitemShut {NoStop}%
\bibitem [{\citenamefont {Beaver}\ and\ \citenamefont
  {Tuck}(1998)}]{beaver1998adjustment}%
  \BibitemOpen
  \bibfield  {author} {\bibinfo {author} {\bibfnamefont {B.}~\bibnamefont
  {Beaver}}\ and\ \bibinfo {author} {\bibfnamefont {B.}~\bibnamefont {Tuck}},\
  }\href@noop {} {\bibfield  {journal} {\bibinfo  {journal} {New Zealand
  Journal of Educational Studies}\ } (\bibinfo {year} {1998})}\BibitemShut
  {NoStop}%
\bibitem [{\citenamefont {Wan}(1999)}]{wan1999learning}%
  \BibitemOpen
  \bibfield  {author} {\bibinfo {author} {\bibfnamefont {G.}~\bibnamefont
  {Wan}},\ }\href@noop {} {\  (\bibinfo {year} {1999})}\BibitemShut {NoStop}%
\bibitem [{\citenamefont {Hu}(2002)}]{hu2002potential}%
  \BibitemOpen
  \bibfield  {author} {\bibinfo {author} {\bibfnamefont {G.}~\bibnamefont
  {Hu}},\ }\href@noop {} {\bibfield  {journal} {\bibinfo  {journal} {Language
  culture and curriculum}\ }\textbf {\bibinfo {volume} {15}},\ \bibinfo {pages}
  {93} (\bibinfo {year} {2002})}\BibitemShut {NoStop}%
\bibitem [{\citenamefont {Wilkinson}\ and\ \citenamefont
  {Olliver-Gray}(2006)}]{wilkinson2006significance}%
  \BibitemOpen
  \bibfield  {author} {\bibinfo {author} {\bibfnamefont {L.}~\bibnamefont
  {Wilkinson}}\ and\ \bibinfo {author} {\bibfnamefont {Y.}~\bibnamefont
  {Olliver-Gray}},\ }\href@noop {} {\bibfield  {journal} {\bibinfo  {journal}
  {Psychologia}\ }\textbf {\bibinfo {volume} {49}},\ \bibinfo {pages} {74}
  (\bibinfo {year} {2006})}\BibitemShut {NoStop}%
\bibitem [{\citenamefont {Chemers}\ \emph {et~al.}(2011)\citenamefont
  {Chemers}, \citenamefont {Zurbriggen}, \citenamefont {Syed}, \citenamefont
  {Goza},\ and\ \citenamefont {Bearman}}]{chemers2011role}%
  \BibitemOpen
  \bibfield  {author} {\bibinfo {author} {\bibfnamefont {M.~M.}\ \bibnamefont
  {Chemers}}, \bibinfo {author} {\bibfnamefont {E.~L.}\ \bibnamefont
  {Zurbriggen}}, \bibinfo {author} {\bibfnamefont {M.}~\bibnamefont {Syed}},
  \bibinfo {author} {\bibfnamefont {B.~K.}\ \bibnamefont {Goza}}, \ and\
  \bibinfo {author} {\bibfnamefont {S.}~\bibnamefont {Bearman}},\ }\href@noop
  {} {\bibfield  {journal} {\bibinfo  {journal} {Journal of Social Issues}\
  }\textbf {\bibinfo {volume} {67}},\ \bibinfo {pages} {469} (\bibinfo {year}
  {2011})}\BibitemShut {NoStop}%
\bibitem [{\citenamefont {Byars-Winston}\ \emph {et~al.}(2015)\citenamefont
  {Byars-Winston}, \citenamefont {Branchaw}, \citenamefont {Pfund},
  \citenamefont {Leverett},\ and\ \citenamefont
  {Newton}}]{byars2015culturally}%
  \BibitemOpen
  \bibfield  {author} {\bibinfo {author} {\bibfnamefont {A.~M.}\ \bibnamefont
  {Byars-Winston}}, \bibinfo {author} {\bibfnamefont {J.}~\bibnamefont
  {Branchaw}}, \bibinfo {author} {\bibfnamefont {C.}~\bibnamefont {Pfund}},
  \bibinfo {author} {\bibfnamefont {P.}~\bibnamefont {Leverett}}, \ and\
  \bibinfo {author} {\bibfnamefont {J.}~\bibnamefont {Newton}},\ }\href@noop {}
  {\bibfield  {journal} {\bibinfo  {journal} {International journal of science
  education}\ }\textbf {\bibinfo {volume} {37}},\ \bibinfo {pages} {2533}
  (\bibinfo {year} {2015})}\BibitemShut {NoStop}%
\bibitem [{\citenamefont {Zohrabi-Alaee}\ and\ \citenamefont
  {Zwickl}(2021)}]{Zohrabi_perc2021}%
  \BibitemOpen
  \bibfield  {author} {\bibinfo {author} {\bibfnamefont {D.}~\bibnamefont
  {Zohrabi-Alaee}}\ and\ \bibinfo {author} {\bibfnamefont {B.~M.}\ \bibnamefont
  {Zwickl}},\ }in\ \href@noop {} {\emph {\bibinfo {booktitle} {Physics
  Education Research Conference 2021}}},\ \bibinfo {series and number} {PER
  Conference}\ (\bibinfo {address} {Virtual Conference},\ \bibinfo {year}
  {2021})\ pp.\ \bibinfo {pages} {480--485}\BibitemShut {NoStop}%
\bibitem [{\citenamefont {Nagda}\ \emph {et~al.}(1998)\citenamefont {Nagda},
  \citenamefont {Gregerman}, \citenamefont {Jonides}, \citenamefont {von
  Hippel},\ and\ \citenamefont {Lerner}}]{nagda_undergraduate_1998}%
  \BibitemOpen
  \bibfield  {author} {\bibinfo {author} {\bibfnamefont {B.~A.}\ \bibnamefont
  {Nagda}}, \bibinfo {author} {\bibfnamefont {S.~R.}\ \bibnamefont
  {Gregerman}}, \bibinfo {author} {\bibfnamefont {J.}~\bibnamefont {Jonides}},
  \bibinfo {author} {\bibfnamefont {W.}~\bibnamefont {von Hippel}}, \ and\
  \bibinfo {author} {\bibfnamefont {J.~S.}\ \bibnamefont {Lerner}},\ }\href
  {\doibase https://doi.org/10.1353/rhe.1998.0016} {\bibfield  {journal}
  {\bibinfo  {journal} {Rev. High. Ed.}\ }\textbf {\bibinfo {volume} {22}},\
  \bibinfo {pages} {55} (\bibinfo {year} {1998})}\BibitemShut {NoStop}%
\bibitem [{\citenamefont {Rodenbusch}\ \emph {et~al.}(2016)\citenamefont
  {Rodenbusch}, \citenamefont {Hernandez}, \citenamefont {Simmons},\ and\
  \citenamefont {Dolan}}]{rodenbusch_early_2016}%
  \BibitemOpen
  \bibfield  {author} {\bibinfo {author} {\bibfnamefont {S.~E.}\ \bibnamefont
  {Rodenbusch}}, \bibinfo {author} {\bibfnamefont {P.~R.}\ \bibnamefont
  {Hernandez}}, \bibinfo {author} {\bibfnamefont {S.~L.}\ \bibnamefont
  {Simmons}}, \ and\ \bibinfo {author} {\bibfnamefont {E.~L.}\ \bibnamefont
  {Dolan}},\ }\href@noop {} {\bibfield  {journal} {\bibinfo  {journal} {CBE
  Life Sci. Educ.}\ }\textbf {\bibinfo {volume} {15}} (\bibinfo {year}
  {2016})}\BibitemShut {NoStop}%
\bibitem [{\citenamefont {Junge}\ \emph {et~al.}(2010)\citenamefont {Junge},
  \citenamefont {Quiñones}, \citenamefont {Kakietek}, \citenamefont
  {Teodorescu},\ and\ \citenamefont {Marsteller}}]{junge_promoting_2010}%
  \BibitemOpen
  \bibfield  {author} {\bibinfo {author} {\bibfnamefont {B.}~\bibnamefont
  {Junge}}, \bibinfo {author} {\bibfnamefont {C.}~\bibnamefont {Quiñones}},
  \bibinfo {author} {\bibfnamefont {J.}~\bibnamefont {Kakietek}}, \bibinfo
  {author} {\bibfnamefont {D.}~\bibnamefont {Teodorescu}}, \ and\ \bibinfo
  {author} {\bibfnamefont {P.}~\bibnamefont {Marsteller}},\ }\href {\doibase
  https://doi.org/10.1187/cbe.09-08-0057} {\bibfield  {journal} {\bibinfo
  {journal} {CBE Life Sci. Educ.}\ }\textbf {\bibinfo {volume} {9}},\ \bibinfo
  {pages} {119} (\bibinfo {year} {2010})}\BibitemShut {NoStop}%
\bibitem [{\citenamefont {Engestr{\"o}m}(2001)}]{Engestrom2001}%
  \BibitemOpen
  \bibfield  {author} {\bibinfo {author} {\bibfnamefont {Y.}~\bibnamefont
  {Engestr{\"o}m}},\ }\href@noop {} {\bibfield  {journal} {\bibinfo  {journal}
  {J. Educ. Work.}\ }\textbf {\bibinfo {volume} {14}},\ \bibinfo {pages} {133}
  (\bibinfo {year} {2001})}\BibitemShut {NoStop}%
\bibitem [{\citenamefont {M.~Hubenthal}\ and\ \citenamefont
  {Frassetto}(2007)}]{2007AGUFMED33B1226H}%
  \BibitemOpen
  \bibfield  {author} {\bibinfo {author} {\bibfnamefont {R.~A.}\ \bibnamefont
  {M.~Hubenthal}, \bibfnamefont {J.~Taber}}\ and\ \bibinfo {author}
  {\bibfnamefont {A.}~\bibnamefont {Frassetto}},\ }in\ \href@noop {} {\emph
  {\bibinfo {booktitle} {Fall Meeting Abstracts}}},\ Vol.\ \bibinfo {volume}
  {2007}\ (\bibinfo {year} {2007})\ pp.\ \bibinfo {pages}
  {33--1226}\BibitemShut {NoStop}%
\bibitem [{\citenamefont {Gonzalez-Ramirez}\ \emph {et~al.}(2021)\citenamefont
  {Gonzalez-Ramirez}, \citenamefont {Mulqueen}, \citenamefont {Zealand},
  \citenamefont {Silverstein}, \citenamefont {Mulqueen},\ and\ \citenamefont
  {BuShell}}]{gonzalez2021emergency}%
  \BibitemOpen
  \bibfield  {author} {\bibinfo {author} {\bibfnamefont {J.}~\bibnamefont
  {Gonzalez-Ramirez}}, \bibinfo {author} {\bibfnamefont {K.}~\bibnamefont
  {Mulqueen}}, \bibinfo {author} {\bibfnamefont {R.}~\bibnamefont {Zealand}},
  \bibinfo {author} {\bibfnamefont {S.}~\bibnamefont {Silverstein}}, \bibinfo
  {author} {\bibfnamefont {C.}~\bibnamefont {Mulqueen}}, \ and\ \bibinfo
  {author} {\bibfnamefont {S.}~\bibnamefont {BuShell}},\ }\href@noop {}
  {\bibfield  {journal} {\bibinfo  {journal} {College Student Journal}\
  }\textbf {\bibinfo {volume} {55}},\ \bibinfo {pages} {29} (\bibinfo {year}
  {2021})}\BibitemShut {NoStop}%
\bibitem [{\citenamefont {Nambiar}(2020)}]{nambiar2020impact}%
  \BibitemOpen
  \bibfield  {author} {\bibinfo {author} {\bibfnamefont {D.}~\bibnamefont
  {Nambiar}},\ }\href@noop {} {\bibfield  {journal} {\bibinfo  {journal} {The
  International Journal of Indian Psychology}\ }\textbf {\bibinfo {volume}
  {8}},\ \bibinfo {pages} {783} (\bibinfo {year} {2020})}\BibitemShut {NoStop}%
\bibitem [{\citenamefont {Zhang}\ \emph {et~al.}(2020)\citenamefont {Zhang},
  \citenamefont {Wang}, \citenamefont {Yang},\ and\ \citenamefont
  {Wang}}]{zhang2020suspending}%
  \BibitemOpen
  \bibfield  {author} {\bibinfo {author} {\bibfnamefont {W.}~\bibnamefont
  {Zhang}}, \bibinfo {author} {\bibfnamefont {Y.}~\bibnamefont {Wang}},
  \bibinfo {author} {\bibfnamefont {L.}~\bibnamefont {Yang}}, \ and\ \bibinfo
  {author} {\bibfnamefont {C.}~\bibnamefont {Wang}},\ }\href@noop {} {\enquote
  {\bibinfo {title} {Suspending classes without stopping learning: China’s
  education emergency management policy in the covid-19 outbreak},}\ }
  (\bibinfo {year} {2020})\BibitemShut {NoStop}%
\end{thebibliography}%
\end{document}